\documentclass[aps,prl,groupaddress,superscriptaddress,reprint,amsmath,amssymb,amsfonts,graphicx,nobibnotes,nofootinbib]{revtex4-2}
\usepackage{graphicx}
\usepackage{physics}
\usepackage{bm}
\usepackage{bbold}
\usepackage{xcolor}
\usepackage{appendix}
\usepackage{tikz}

\begin{document}

\title{Resolving the backbone of multiscale interactions through mutual information rates}

\author{Alberto Raiola}
\affiliation{Department of Physics and Astronomy ``G. Galilei'', University of Padova, 35131 Padova, Italy}
\affiliation{INFN, Padova Section, 35131 Padova, Italy}

\author{Giorgio Nicoletti}
\email{D.M.B. and G.N. contributed equally to this work}
\affiliation{Quantitative Life Sciences section, The Abdus Salam International Center for Theoretical Physics (ICTP), Trieste, Italy}
\affiliation{Theoretical Science Visiting Program (TSVP), Okinawa Institute of Science and Technology, Onna, Okinawa, Japan}

\author{Daniel M. Busiello}
\email{D.M.B. and G.N. contributed equally to this work}
\affiliation{Department of Physics and Astronomy ``G. Galilei'', University of Padova, 35131 Padova, Italy}
\affiliation{Theoretical Science Visiting Program (TSVP), Okinawa Institute of Science and Technology, Onna, Okinawa, Japan}
\affiliation{Max Planck Institute for the Physics of Complex Systems, 01187 Dresden, Germany}

\begin{abstract}
\noindent Natural systems are often composed of processes evolving on multiple timescales and coupled through complex interaction networks. Inferring such networks is especially challenging when timescales are widely separated, since resolving the fastest processes while accurately sampling the slowest ones is generally prohibitive. In this work, we propose a framework to reconstruct a multiscale interaction tree by combining a rescaled version of the mutual information rate, its conditional counterpart, and the ordering of timescales. We derive an exact theory for stationary Gaussian processes, finding that these measures identify directly connected processes, with the asymptotic mutual information rate distinguishing fast-to-slow from slow-to-fast couplings. We show that this reconstruction can be achieved from the spectral matrix and illustrate its robustness to finite trajectory lengths, moderate timescale separation, and weak nonlinearities. Our work paves the way for exploiting the presence of multiple timescales in real-world systems as a source of structural identifiability rather than an obstacle to network inference.
\end{abstract}

\maketitle

\noindent Natural systems are often characterized by interconnected processes operating across multiple temporal scales, with fast and slow variables coupled through complex multiscale interaction networks \cite{hastings2010timescales, de2023more, schaffer2021mapping}. This multiscale organization plays a central role in shaping emergent behavior. Among the many examples spanning physics and biology, dissipative and functional pathways can be selected out of equilibrium by slow-switching drivings \cite{brandman2005interlinked, brandman2008feedback, dass2021equilibrium, bilancioni2025gears}, adaptation and habituation in signaling networks are mediated by slow feedback mechanisms \cite{eckert2024biochemically, nicoletti2025optimal, yi2000robust, aoki2019universal, lan2012energy, matsumoto2018role}, DNA methylation aging on long timescales is controlled by fast molecular kinetics \cite{ciarchi2025temporal, pridham2024dynamical}, rich dynamical features in chemical networks can emerge through the interplay of fast reactions and regulatory processes \cite{dack2026recurrent,mengel2010modeling}, and computation in neural populations is shaped by heterogeneous dynamical timescale and plasticity mechanisms \cite{gao2020neuronal, golesorkhi2021brain, o2023nonlinear, zeraati2026neural}. Most of these phenomena rely on interactions that bridge different temporal scales and couple fast and slow degrees of freedom (dofs).

Information theory provides several tools to quantify such couplings at a statistical level \cite{horowitz2014thermodynamics}. For any two dynamical variables $X$ and $Y$, their dependence at time $t$ is quantified by the mutual information,
\begin{equation}
    I(X_t;Y_t) = \int dx dy ~p_{X_tY_t}(x,y) \log\frac{p_{X_tY_t}(x,y)}{p_{X_t}(x) p_{Y_t}(y)} \;,
    \label{eq:MI}
\end{equation}
which is the Kullback-Leibler divergence between the joint distribution $p_{X_tY_t}$ and the product of the marginals $p_{X_t} p_{Y_t}$. Examining how mutual information propagates across timescales \cite{nicoletti2024information, nicoletti2024gaussian} reveals that only a subset of dynamical couplings is \emph{functional}, i.e., able to generate statistical dependence between fast and slow variables. In particular, while \emph{feedback} interactions from slow to fast dofs generate information, \emph{direct} interactions from fast to slow dofs can only propagate it on longer timescales. This set of rules imposes important constraints on the temporal architecture of information-processing systems \cite{nicoletti2025stochastic}. Nevertheless, only some interaction paths across different timescales contribute to the mutual information between variables, and thus $I(X_t;Y_t)$ cannot easily be exploited to infer the structure of the underlying multiscale interaction network.

Furthermore, $I(X_t;Y_t)$ only accounts for the information shared, on average, by the instantaneous values of the two processes, neglecting statistical dependencies encoded in their temporal evolution. To overcome this limitation, one can consider finite trajectories as observables, $\bm{X}_T = \{X_t\}_{0\leq t\leq T}$ and $\bm{Y}_T = \{Y_t\}_{0\leq t\leq T}$, and compute $I(\bm{X}_T;\bm{Y}_T)$ again via Eq.~\eqref{eq:MI} with the corresponding path distributions. For stationary processes with suitable regularity conditions \cite{kolmogorov1956shannon, komaee2020mutual}, $I(\bm{X}_T;\bm{Y}_T)$ grows asymptotically linearly with the observation time $T$, so that one can define the mutual information rate as
\begin{equation}
    R_{XY} = \lim_{T\to \infty} \frac{1}{T} I(\bm{X}_T;\bm{Y}_T) \; .
\end{equation}
Analogously, the conditional mutual information rate $R_{XY|Z}$ quantifies the information shared by two trajectories conditioned on the trajectories $\bm{Z}_T$ of other processes.

While the mutual information rate has been extensively studied in the context of chemical signaling networks \cite{tostevin2009mutual, moor2023dynamic, reinhardt2023path,tkavcik2025information, deronde2010effect, deronde2012feedforward, mattingly2021escherichia}, its interplay with the timescales governing the evolution of the underlying processes remains unexplored.
In this work, we investigate this idea by first focusing on multiscale Gaussian processes, a setting that is both analytically tractable and known to approximate the behavior of large chemical systems \cite{tostevin2009mutual, tostevin2010mutual, moor2026state}. We show that, by combining ordinary and conditional mutual information rates, we can exactly identify the links of a multiscale interaction tree when the timescales of the underlying processes are sufficiently separated. Crucially, this reconstruction can be performed directly from the power spectra of the observed trajectories, without fitting a multivariate dynamical model \cite{frishman2020learning}, and shows only limited degeneracy for networks with interaction loops. Finally, we test the robustness of this spectral approach against finite trajectory lengths, finite timescale separation, and nonlinear dynamics.

It is instructive to start from the simple example of two dofs, a sender $x_\mathrm{S}$ and a receiver $x_\mathrm{R}$, where $x_\mathrm{S}$ influences the dynamics of $x_\mathrm{R}$. We denote this scenario by $x_\mathrm{S} \Rightarrow x_\mathrm{R}$ hereafter. The evolution of the sender and receiver dofs is characterized by two distinct timescales, $\tau_\mathrm{S}$ and $\tau_\mathrm{R}$ respectively:
\begin{equation}
    \begin{gathered}
    \label{eq:2_langevin}
    \tau_\mathrm{S} \frac{d}{dt}x_\mathrm{S} = - x_\mathrm{S} + \sqrt{2 D_\mathrm{S} \tau_\mathrm{S}} \, \xi_\mathrm{S} \\
    \tau_\mathrm{R} \frac{d}{dt}x_\mathrm{R} = - x_\mathrm{R} + a \,x_\mathrm{S} + \sqrt{2 D_\mathrm{R} \tau_\mathrm{R}} \, \xi_\mathrm{R} \;,
\end{gathered}
\end{equation}
where $a$ is the coupling strength, $D_i$ are the diffusion coefficients, and $\xi_i$ are Gaussian white noises with zero mean and unit variance, for $i \in \{\mathrm{S},\mathrm{R}\}$. The system described by Eq.~\eqref{eq:2_langevin} has, by construction, a single timescale for each process and, at equilibrium, is compatible with the fluctuation-dissipation relation \cite{di2023variance, nicoletti2026balancing}. The linear interaction in Eq.~\eqref{eq:2_langevin} makes it possible to write the mutual information rate as
\begin{equation}
\label{eq:mutual_rate_spectrum}
    R_{\mathrm{S}\mathrm{R}} = -\frac{1}{4\pi} \int_{-\infty}^{+\infty} d\omega ~\log \left(1 - \phi_{\mathrm{S}\mathrm{R}}(\omega)\right)
\end{equation}
where $R_{\mathrm{S}\mathrm{R}}$ is a shorthand notation for $R_{X_\mathrm{S}X_\mathrm{R}}$, and
\begin{equation}
    \phi_{\mathrm{S}\mathrm{R}}(\omega) = \frac{|S_{\mathrm{S}\mathrm{R}}(\omega)|^2}{S_{\mathrm{S}\mathrm{S}}(\omega)S_{\mathrm{R}\mathrm{R}}(\omega)}
\end{equation}
is the coherence, with $S_{ij}(\omega)$ the $ij$-th element of the power spectrum, i.e., the Fourier transform of the correlation function $C_{ij}(t)$. We note that Eq.~\eqref{eq:mutual_rate_spectrum} holds for any Gaussian process \cite{tostevin2009mutual,munakata2006stochastic}. A well-known result for Gaussian channels is that, if $x_\mathrm{S} \Rightarrow x_\mathrm{R}$, the receiver acts as a low-pass filter, attenuating fluctuations from the sender at frequencies above $\tau_\mathrm{R}^{-1}$ \cite{tostevin2009mutual}.

\begin{figure}
    \centering
    \includegraphics[width=\columnwidth]{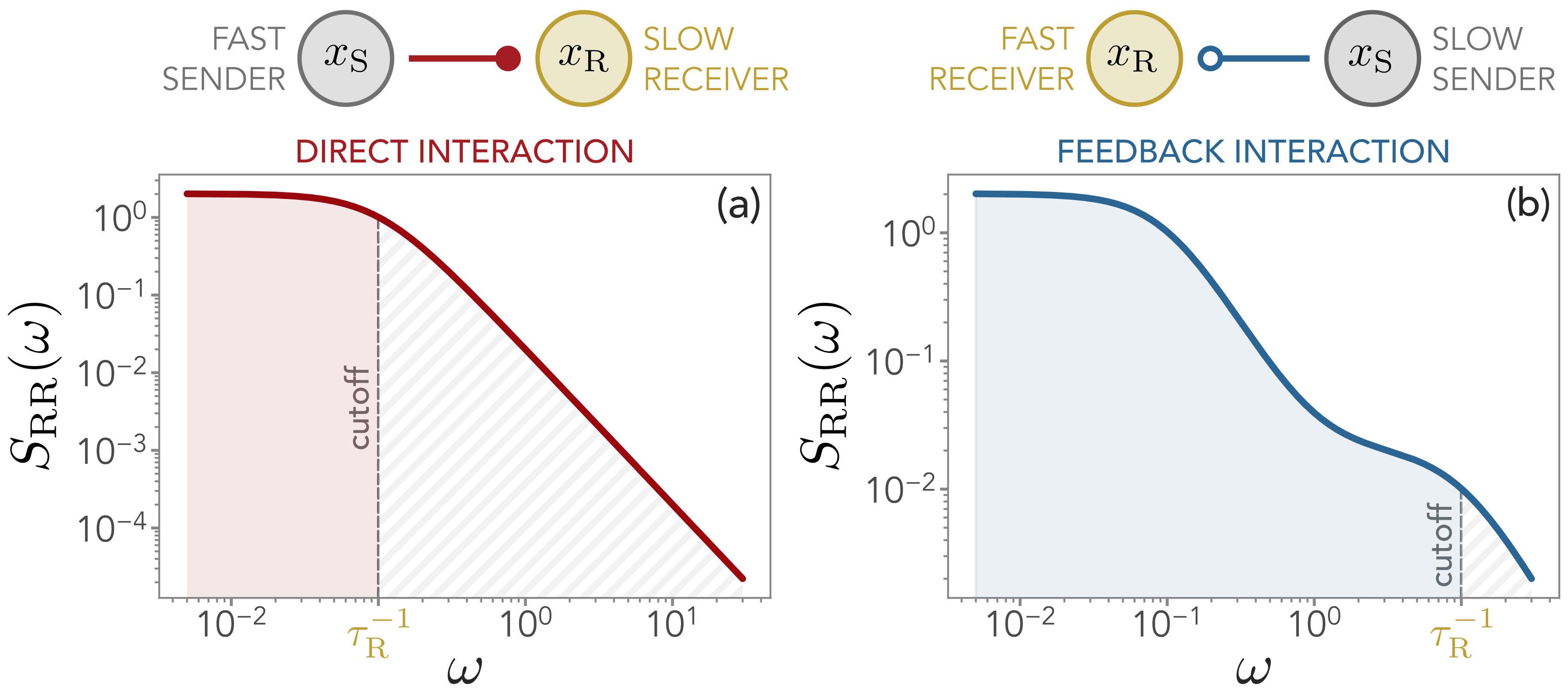}
    \caption{A sender dof ($x_\mathrm{S}$) influences the dynamics of a receiver ($x_\mathrm{R}$). We show the receiver component of the power spectrum, $S_{\mathrm{RR}}$. The two dofs evolve on different timescales, $\tau_\mathrm{S}$ and $\tau_\mathrm{R}$, with $\tau_\mathrm{R}^{-1}$ setting the cutoff frequency of the receiver. With a direct interaction ($\tau_\mathrm{S} \ll \tau_\mathrm{R}$), the receiver behaves as a low-pass filter, and the rescaled mutual information rate $\hat{R}_{\mathrm{S}\mathrm{R}}$ remains finite. In the opposite limit of a feedback interaction ($\tau_\mathrm{S} \gg \tau_\mathrm{R}$), the cutoff becomes large and the receiver is able to follow the sender's fluctuations over a broader frequency range, making the mutual information rate diverge in the timescale separation limit. In this Figure, $D_\mathrm{R} = D_\mathrm{S} = 1$ and $a = 1$. The fast and slow dofs have timescales $\tau = 10^{-1}$ and $\tau = 10^1$ respectively.}
    \label{fig:sketch}
\end{figure}

In a multiscale scenario, $x_\mathrm{S}$ and $x_\mathrm{R}$ evolve on very different timescales. Hence, to observe the dynamics of both processes, and therefore compute $R_{\mathrm{S}\mathrm{R}}$, the observation time must span the timescale of the slowest process. This amounts to rescaling the time as $t \to t' = t /\max\{\tau_\mathrm{S},\tau_\mathrm{R}\}$ and the frequency as $\omega \to \omega' = \omega\max\{\tau_\mathrm{S},\tau_\mathrm{R}\}$. The corresponding rescaled mutual information rate is denoted by $\hat{R}_{\mathrm{S}\mathrm{R}}$. Timescale separation is obtained by keeping the observation timescale finite and letting $\min(\tau_S,\tau_R) \to 0$ \cite{bo2017multiple}. Figure \ref{fig:sketch} sketches the two possible scenarios:
\begin{itemize}
    \item[(i)] When the sender $x_\mathrm{S}$ is faster than the receiver $x_\mathrm{R}$, we have a direct interaction. The slow receiver strongly attenuates the sender's fast fluctuations above its finite cutoff $\tau_\mathrm{R}^{-1}$, and the rescaled mutual information rate $\hat{R}_{\mathrm{SR}}$ remains finite.
    \item[(ii)] When the sender is slower than the receiver, we have a feedback interaction. The cutoff $\tau_\mathrm{R}^{-1}$ diverges in the timescale-separation limit, allowing the receiver to follow the sender's fluctuations over a growing frequency range. In this limit, $\hat{R}_{\mathrm{SR}}\to\infty$.
\end{itemize}
This result is the starting point of our work. Its analytical derivation is given in the Supplemental Material (SM) \cite{supplemental_material}.

\begin{figure*}
    \centering
    \includegraphics[width=\textwidth]{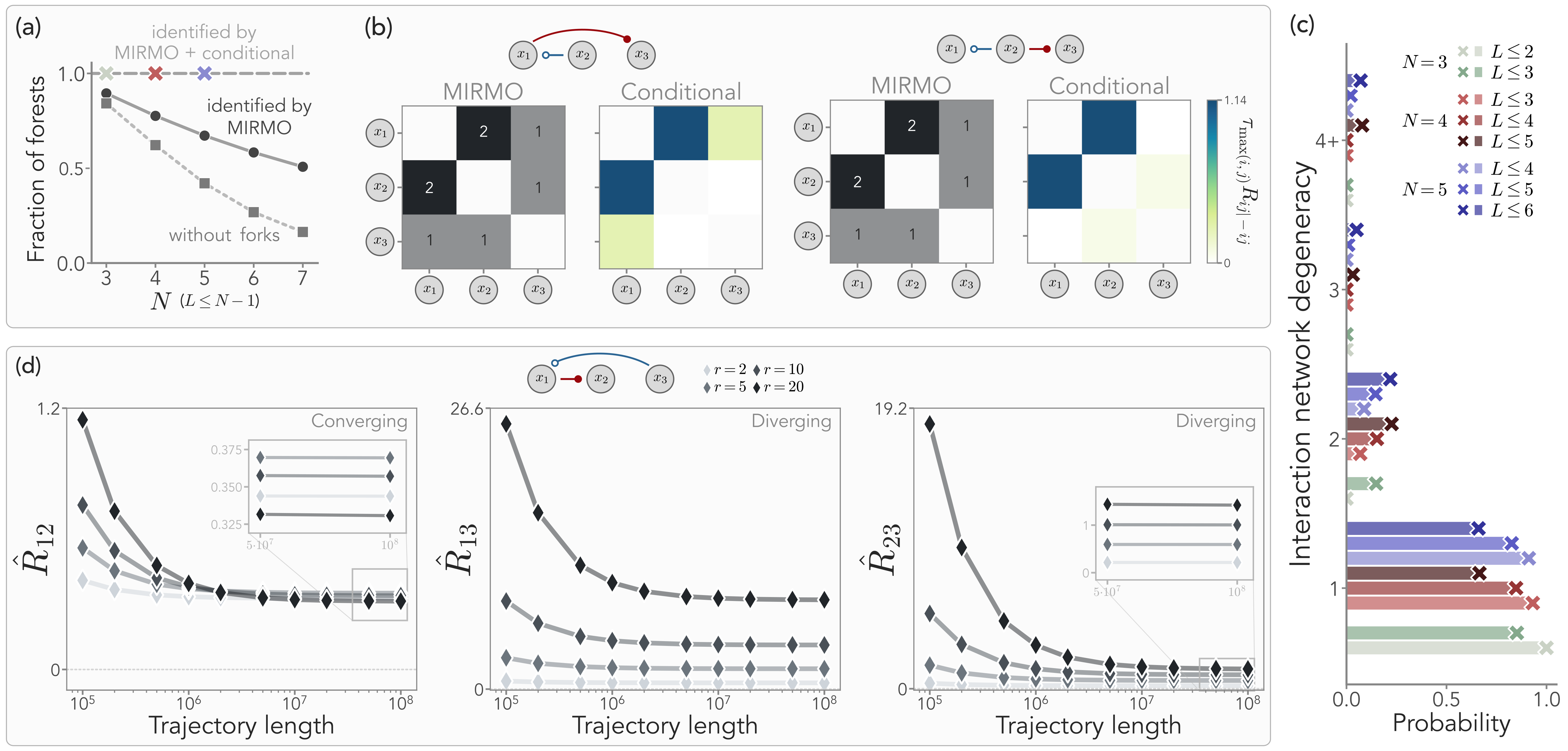}
    \caption{(a) Fraction of forests uniquely identified by the MIRMO alone and by combining it with the zero entries of the conditional mutual information rate, as a function of the number of nodes $N$, with $L\leq N-1$ links. The dashed gray curve with square markers shows the overall fraction of forests without forks. (b) The chain $x_2\Rightarrow x_1\Rightarrow x_3$ (left) and the fork $x_2\Rightarrow x_1$, $x_2\Rightarrow x_3$ (right) share the same MIRMO labels, $(\mathcal L_{12},\mathcal L_{23},\mathcal L_{13})=(2,1,1)$. However, their conditional rates, plotted as $\tau_{\max(i,j)}R_{ij|-ij}$, distinguish the two architectures. (c) Distribution of the interaction-network degeneracy for $N=3,4,5$ at different maximum numbers of links. This plot was obtained by combining the MIRMO in the timescale separation limit with the zero entries of the conditional mutual information rate at finite $r$. (d) Rescaled mutual information rates $\hat R_{ij}$ estimated from trajectories of $x_3\Rightarrow x_1\Rightarrow x_2$ as a function of trajectory length and at different values of $r$. When $r\to\infty$, $\hat R_{12}$ converges to a finite value, while $\hat R_{13}$ and $\hat R_{23}$ diverge. In panel (d), all diffusion coefficients are set to $D_i=1$, the nonzero coupling strengths are all $a_{ij}=1$, and the integration step is $dt=0.1$.}
    \label{fig:linear}
\end{figure*}

Consider now a set $\mathcal{X} = \{x_1, x_2, \dots, x_N\}$ of $N$ coupled Gaussian processes, ordered from the fastest to the slowest timescale, $\tau_1 < \tau_2 < \dots < \tau_N$. We set $\tau_{i+1}/\tau_i = r$, so that timescale separation is obtained for $r\to \infty$. We first focus on tree-like interaction networks with $N-1$ interactions and no disconnected dofs. We refer to this structure as the \emph{backbone} of multiscale interactions. For any two processes in $\mathcal{X}$, $x_i$ and $x_j$, we can rescale their evolution by the slowest timescale between $\tau_i$ and $\tau_j$ and define the corresponding rescaled mutual information rate, $\hat{R}_{ij}$ (see Eq.~\eqref{eq:mutual_rate_spectrum}). As shown above, if $x_i$ and $x_j$ are coupled, $\hat{R}_{ij}$ is nonzero and can either remain finite or diverge in the timescale separation limit, depending on whether they share a direct or a feedback interaction, respectively. To summarize this behavior, we introduce a labeling operator $\mathcal{L}_{ij}$ that takes the value $2$ when $\hat{R}_{ij}$ diverges, $1$ when it is finite, and $0$ when it vanishes. The first important result, discussed in the End Matter, is that, in an interaction tree, the labels associated with a direct or feedback interaction are not affected by the surrounding topology.

Moreover, $x_i$ and $x_j$ may also be coupled indirectly through other dofs so that $\hat{R}_{ij}$ may be nonzero even when $x_i$ and $x_j$ share a common \textit{ancestor}, i.e., they are driven by the same dof. However, in the timescale separation limit, the labels of these indirect interactions can be determined through additional simple rules. Specifically, if $x_i$ and $x_j$ are connected by a feedback interaction followed by a direct one, if the latter is enclosed in the former, its label is $2$. Conversely, the label reduces by one (at most to $0$) for each direct interaction extending beyond a feedback one. A detailed discussion of these cases is presented in the SM \cite{supplemental_material}. Intuitively, these rules reflect the interplay between timescale ordering and the range of frequencies that effectively transmit information between different dofs. All these results extend to forests, i.e., collections of disconnected trees.

According to the rules above, by collecting the labels of $\hat{R}_{ij}$ for all pairs of processes, we can construct the \textbf{M}utual \textbf{I}nformation \textbf{R}ate matrix for \textbf{M}ultiscale \textbf{O}bservables (MIRMO), a symmetric $N \times N$ matrix with entries in $\{0,1,2\}$ that summarizes how information propagates across timescales. In Figure~\ref{fig:linear}a, we show that the fraction of forests identified by the MIRMO does not reach $1$ and worsens as the network size increases. In Figure~\ref{fig:linear}b, we show one case in which the MIRMO is identical for two different three-node structures, with one of them being a fork, i.e., a dof with more than one outgoing interaction. Although every identifiability issue involves a fork, not all forks create one (see Fig.~\ref{fig:linear}a).

To resolve this problem, we introduce the rescaled mutual information rate conditioned on the full trajectories of all dofs but $x_i$ and $x_j$, denoted by $\hat{R}_{ij|-ij}$. This quantity explicitly retains information on whether two different dofs appear in the same dynamical equation (see End Matter). It is defined as in Eq.~\eqref{eq:mutual_rate_spectrum} with the coherence replaced by
\begin{equation}
    \phi_{ij|-ij}(\omega') = \frac{|P_{ij}(\omega')|^2}{P_{ii}(\omega')P_{jj}(\omega')} \;,
\end{equation}
where $P = S^{-1}$ is the inverse spectral matrix \cite{Dahlhaus2000}. Unlike MIRMO, the matrix with entries $\hat{R}_{ij|-ij}$, which we call \textit{the conditional matrix}, distinguishes the two scenarios (see Figure \ref{fig:linear}b). More importantly, combining the MIRMO and the conditional matrix, we identify all forests regardless of network size (see Fig.~\ref{fig:linear}a). A demonstration that every pair constituted by a MIRMO and a conditional matrix corresponds to one and only one set of $N-1$ interactions connecting $N$ processes is presented in the End Matter.

The bijection holds for tree-like and forest structures. For $L>N-1$, however, the interaction network can contain loops, so that the inference is no longer unique and the reconstructed networks may exhibit some degeneracy.
In Figure \ref{fig:linear}c, we consider graphs with increasing numbers of links and different numbers of nodes, and show the number of interaction networks compatible with a single pair of a MIRMO and a conditional matrix. As expected, tree-like architectures are always reconstructed exactly. For $L > N-1$, the degeneracy increases with the number of additional links, although a non-negligible fraction of networks is still uniquely identified. Note that networks with $L \leq N-1$ can still generate loops if some dofs are disconnected.

\begin{figure}
    \centering
    \includegraphics[width=\columnwidth]{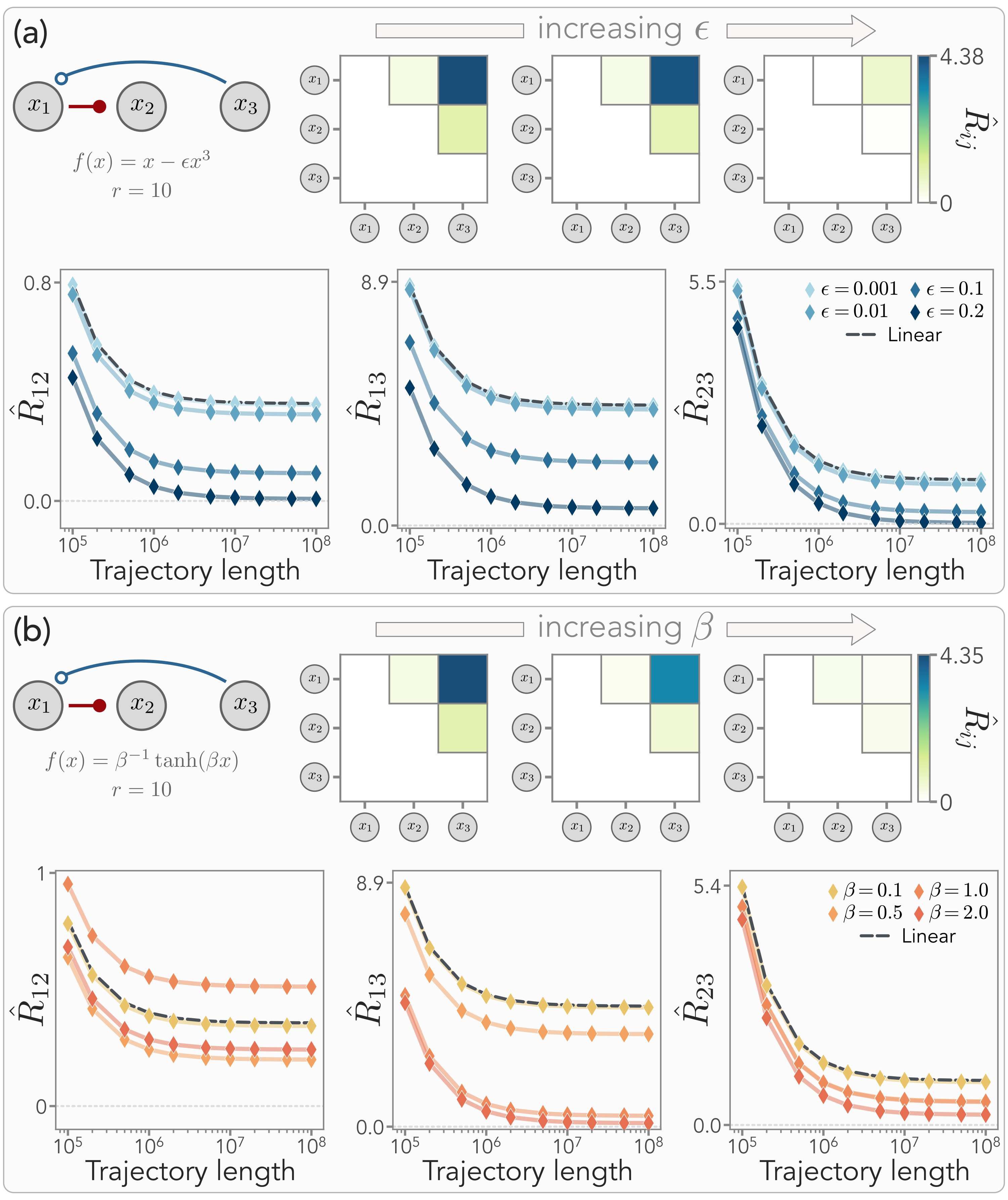}
    \caption{(a) To assess whether the properties of the mutual information rate for a linear system remain qualitatively robust with nonlinear interactions, we consider a cubic nonlinearity for the tree network $3 \Rightarrow 1 \Rightarrow 2$ at fixed timescale separation $r = 10$. As the strength of the nonlinear term $\epsilon$ increases, the entries $\hat{R}_{ij}$ shift towards zero, but the overall structure of the matrix remains consistent with the linear case (dashed lines) for sufficiently small $\epsilon$. (b) Same, but for a saturating nonlinearity $f(x) = \beta^{-1}\tanh(\beta x)$. In this case, the effect of the nonlinearity strength $\beta$ becomes more evident at large $\beta$, with the ranking of the elements $\hat{R}_{ij}$ shifting due to the saturating effect. Nevertheless, the Gaussian approximation remains qualitatively valid for small $\beta$. In this Figure, all diffusion coefficients are set to $D_i = 1$, and the integration step for numerical simulations is $dt = 0.1$.}
    \label{fig:nonlinear}
\end{figure}

One of the main advantages of the proposed approach is that both the MIRMO and the conditional matrix can be estimated directly from stochastic trajectories through their power spectra. In particular, timescale separation is needed only to assign labels, i.e., to establish the direction of the inferred interactions. In Figure \ref{fig:linear}d, we assess the accuracy of this reconstruction for increasing trajectory lengths and decreasing timescale separation. For a representative architecture, we find that the expected labels can be recovered even without strong timescale separation when sufficiently long trajectories are available. More generally, real data typically do not exhibit strong timescale separation. In this case, as shown in the End Matter, it is always possible to reconstruct the undirected underlying interaction tree.

We then study numerically whether these spectral signatures remain informative when the dynamics is not exactly Gaussian,
\begin{equation}
    \tau_i \,\frac{dx_i}{dt} = -x_i + \sum_{j \ne i} a_{ij} f(x_j) + \sqrt{2 D_i \tau_i} \xi_i
\end{equation}
for $i = 1, \dots, N$. For these trajectories, the coherence integral in Eq.~\eqref{eq:mutual_rate_spectrum} gives a Gaussian spectral approximation of the mutual information rate. For simplicity, we take $a_{ij} = 1$ if $x_j \Rightarrow x_i$, and $0$ otherwise. In Figure \ref{fig:nonlinear}, we consider the tree-like network $3 \Rightarrow 1 \Rightarrow 2$ at fixed timescale separation $r = 10$, for which, in the linear case and in the timescale separation limit, $\mathcal{L}_{12} = 1$ and $\mathcal{L}_{13} = \mathcal{L}_{23} = 2$, and study two qualitatively different nonlinearities. Importantly, when only finite trajectories are accessible, the main difficulty is to distinguish a large but finite $\hat{R}_{ij}$ ($\mathcal{L}_{ij}=1$) from a diverging one ($\mathcal{L}_{ij} = 2$).
Therefore, as a practical approach, since labels cannot be assigned to a single finite estimate of $\hat{R}_{ij}$, we test whether the entries expected to diverge remain clearly separated from those expected to converge. In Figure \ref{fig:nonlinear}a, we show that a cubic nonlinearity $f(x_j) = x_j - \epsilon x_j^3$ shifts all entries $\hat{R}_{ij}$ towards zero, but preserves this separation, so that the spectral signatures remain similar to those of the linear case for sufficiently small $\epsilon$. In Figure \ref{fig:nonlinear}b, we consider instead a saturating nonlinearity $f(x_j) = \beta^{-1}\tanh(\beta x_j)$, which has a stronger effect. As the steepness $\beta$ grows, the entries of the mutual information rate are progressively reshuffled, and the separation is eventually lost. Hence, the Gaussian approximation underlying our approach remains reliable for weak nonlinearities, with the threshold set by how strongly the nonlinearity reshapes the effective couplings between processes. A systematic characterization of this threshold, and of the role of the network size, is left for future work.

In summary, we have shown that a properly labeled and rescaled mutual information rate between pairs of processes evolving on separated timescales contains clear signatures of the architecture of the underlying interaction network. By combining the ordering of characteristic timescales with the asymptotic scaling of the mutual information rate and the conditional one, we recover this network exactly when it exhibits a tree-like structure -- i.e., each pair of processes is connected by a unique pathway. Crucially, our approach only requires estimating the spectral matrix from stochastic trajectories of the system rather than fitting a multivariate dynamical model. Beyond trees, the same spectral signatures still identify the interaction backbone up to a limited degeneracy among compatible network architectures.

The scope of our approach differs from that of causal discovery methods. Granger-causality \cite{shojaie2022granger, faes2017multiscale}, transfer-entropy \cite{schreiber2000measuring, barnett2009granger, runge2012escaping}, and conditional-independence approaches \cite{runge2019detecting} aim to identify directed interactions in broad classes of time-series models \cite{ren2010noise, baldovin2020understanding, auconi2017causal}. However, strong timescale separation makes inference particularly challenging, since trajectories must resolve the fastest dynamics while spanning sufficiently long observation times to capture the slowest ones. Furthermore, information-theoretic measures may assign spurious links between indirectly connected variables, or fail to capture mediated interactions, especially when these act across different timescales \cite{hlavackova2007causality, smirnov2013spurious, james2016information, runge2018causal, del2024robust}. Here, instead, combining ordinary and conditional rates and exploiting timescale ordering allows us to separate such dependence from actual directional links in a tree, beyond the conditional independence structure, as illustrated in the End Matter.

Although the present framework identifies the backbone of multiscale interactions in the form of a tree network, two distinct sources of error should be highlighted. Reconstruction may fail because of finite-sample estimation of mutual information rates, or because the system is too far from the timescale separation regime, making the relevant spectral signatures difficult to resolve. It will therefore be important to characterize how the interplay between intrinsic timescales and observation times determines which features of the interaction network remain identifiable. The remaining degeneracy beyond trees may be tackled in future work by combining mutual information rates evaluated under different conditioning sets. In particular, optimizing such combinations and their relative weights may improve discrimination between compatible network architectures. These extensions would bridge the gap between the temporal structure of multiscale processes and the underlying architecture of their interactions, making timescale hierarchies a useful resource for understanding the organization of complex biological and physical systems.

\begin{acknowledgments}
\textit{Acknowledgments} -- A.R. acknowledges CloudVeneto for providing computing and storage resources. D.M.B. is funded by the STARS@UNIPD program through the project ``ActiveInfo''. Part of this research was conducted during the visit of D.M.B. and G.N. at the Okinawa Institute of Science and Technology (OIST) through the Theoretical Sciences Visiting Program (TSVP).
\end{acknowledgments}

\bibliography{paper_bib}

@article{hastings2010timescales,
  title={Timescales, dynamics, and ecological understanding},
  author={Hastings, Alan},
  journal={Ecology},
  volume={91},
  number={12},
  pages={3471--3480},
  year={2010},
  publisher={Wiley Online Library}
}

@article{dass2021equilibrium,
  title={Equilibrium and non-equilibrium furanose selection in the ribose isomerisation network},
  author={Dass, Avinash Vicholous and Georgelin, Thomas and Westall, Frances and Foucher, Fr{\'e}d{\'e}ric and De Los Rios, Paolo and Busiello, Daniel Maria and Liang, Shiling and Piazza, Francesco},
  journal={Nature communications},
  volume={12},
  number={1},
  pages={2749},
  year={2021},
  publisher={Nature Publishing Group UK London}
}

@article{ciarchi2025temporal,
  title={A temporal hierarchy determines epigenetic aging},
  author={Ciarchi, Matteo and Simons, Benjamin D and Rulands, Steffen},
  journal={bioRxiv},
  pages={2025--08},
  year={2025},
  publisher={Cold Spring Harbor Laboratory}
}

@article{yi2000robust,
  title={Robust perfect adaptation in bacterial chemotaxis through integral feedback control},
  author={Yi, Tau-Mu and Huang, Yun and Simon, Melvin I and Doyle, John},
  journal={Proceedings of the National Academy of Sciences},
  volume={97},
  number={9},
  pages={4649--4653},
  year={2000},
  publisher={The National Academy of Sciences}
}

@article{kolmogorov1956shannon,
  title={On the Shannon theory of information transmission in the case of continuous signals},
  author={Kolmogorov, Andrey},
  journal={IRE Transactions on Information Theory},
  volume={2},
  number={4},
  pages={102--108},
  year={1956},
  publisher={IEEE}
}

@article{komaee2020mutual,
  title={Mutual information rate between stationary Gaussian processes},
  author={Komaee, Arash},
  journal={Results in Applied Mathematics},
  volume={7},
  pages={100107},
  year={2020},
  publisher={Elsevier}
}

@article{tostevin2010mutual,
  title={Mutual information in time-varying biochemical systems},
  author={Tostevin, Filipe and Ten Wolde, Pieter Rein},
  journal={Physical Review E—Statistical, Nonlinear, and Soft Matter Physics},
  volume={81},
  number={6},
  pages={061917},
  year={2010},
  publisher={APS}
}

@article{moor2023dynamic,
  title={Dynamic information transfer in stochastic biochemical networks},
  author={Moor, Anne-Lena and Zechner, Christoph},
  journal={Physical Review Research},
  volume={5},
  number={1},
  pages={013032},
  year={2023},
  publisher={APS}
}

@article{reinhardt2023path,
  title={Path weight sampling: Exact Monte Carlo computation of the mutual information between stochastic trajectories},
  author={Reinhardt, Manuel and Tka{\v{c}}ik, Ga{\v{s}}per and Ten Wolde, Pieter Rein},
  journal={Physical Review X},
  volume={13},
  number={4},
  pages={041017},
  year={2023},
  publisher={APS}
}

@article{moor2026state,
  title={State-versus reaction-based information processing in biochemical networks},
  author={Moor, Anne-Lena and Tjalma, Age and Reinhardt, Manuel and Wolde, Pieter Rein ten and Zechner, Christoph},
  journal={Physical Review Letters},
  volume={136},
  number={24},
  pages={248401},
  year={2026},
  publisher={APS}
}

@article{munakata2006stochastic,
  title={Stochastic resonance in the FitzHugh-Nagumo model from a dynamic mutual information point of view},
  author={Munakata, T and Kamiyabu, M},
  journal={The European Physical Journal B-Condensed Matter and Complex Systems},
  volume={53},
  number={2},
  pages={239--243},
  year={2006},
  publisher={Springer}
}

@article{tkavcik2025information,
  title={Information processing in biochemical networks},
  author={Tka{\v{c}}ik, Ga{\v{s}}per and Wolde, Pieter Rein ten},
  journal={Annual review of biophysics},
  volume={54},
  number={1},
  pages={249--274},
  year={2025},
  publisher={Annual Reviews}
}

@article{runge2019detecting,
  title={Detecting and quantifying causal associations in large nonlinear time series datasets},
  author={Runge, Jakob and Nowack, Peer and Kretschmer, Marlene and Flaxman, Seth and Sejdinovic, Dino},
  journal={Science advances},
  volume={5},
  number={11},
  pages={eaau4996},
  year={2019},
  publisher={American Association for the Advancement of Science}
}

@article{shojaie2022granger,
  title={Granger causality: A review and recent advances},
  author={Shojaie, Ali and Fox, Emily B},
  journal={Annual review of statistics and its application},
  volume={9},
  pages={289--319},
  year={2022},
  publisher={Annual Reviews}
}

@article{faes2017multiscale,
  title = {Multiscale Granger causality},
  author = {Faes, Luca and Nollo, Giandomenico and Stramaglia, Sebastiano and Marinazzo, Daniele},
  journal = {Phys. Rev. E},
  volume = {96},
  issue = {4},
  pages = {042150},
  numpages = {7},
  year = {2017},
  month = {Oct},
  publisher = {American Physical Society},
  doi = {10.1103/PhysRevE.96.042150},
  url = {https://link.aps.org/doi/10.1103/PhysRevE.96.042150}
}

@article{barnett2009granger,
  title={Granger causality and transfer entropy are equivalent for Gaussian variables},
  author={Barnett, Lionel and Barrett, Adam B and Seth, Anil K},
  journal={Physical review letters},
  volume={103},
  number={23},
  pages={238701},
  year={2009},
  publisher={APS}
}

@article{del2024robust,
  title={Robust inference of causality in high-dimensional dynamical processes from the information imbalance of distance ranks},
  author={Del Tatto, Vittorio and Fortunato, Gianfranco and Bueti, Domenica and Laio, Alessandro},
  journal={Proceedings of the National Academy of Sciences},
  volume={121},
  number={19},
  pages={e2317256121},
  year={2024},
  publisher={National Academy of Sciences}
}

@article{dack2026recurrent,
  title={Recurrent neural chemical reaction networks that approximate arbitrary dynamics},
  author={Dack, Alexander and Qureshi, Benjamin and Ouldridge, Thomas E and Plesa, Tomislav},
  journal={Cell Systems},
  volume={17},
  number={5},
  year={2026},
  publisher={Elsevier}
}

@article{mengel2010modeling,
  title={Modeling oscillatory control in NF-$\kappa$B, p53 and Wnt signaling},
  author={Mengel, Benedicte and Hunziker, Alexander and Pedersen, Lykke and Trusina, Ala and Jensen, Mogens H and Krishna, Sandeep},
  journal={Current opinion in genetics \& development},
  volume={20},
  number={6},
  pages={656--664},
  year={2010},
  publisher={Elsevier}
}

@article{bilancioni2025gears,
  title={Gears in chemical reaction networks for optimizing energy transduction efficiency},
  author={Bilancioni, Massimo and Esposito, Massimiliano},
  journal={Nature Communications},
  volume={16},
  number={1},
  pages={5765},
  year={2025},
  publisher={Nature Publishing Group UK, London}
}

@article{schreiber2000measuring,
  title={Measuring information transfer},
  author={Schreiber, Thomas},
  journal={Physical review letters},
  volume={85},
  number={2},
  pages={461},
  year={2000},
  publisher={APS}
}

@article{hlavackova2007causality,
  author  = {Hlav{\'a}{\v{c}}kov{\'a}-Schindler, Katerina and Palu{\v{s}}, Milan and Vejmelka, Martin and Bhattacharya, Joydeep},
  title   = {Causality detection based on information-theoretic approaches in time series analysis},
  journal = {Physics Reports},
  volume  = {441},
  number  = {1},
  pages   = {1--46},
  year    = {2007},
  doi     = {10.1016/j.physrep.2006.12.004}
}

@article{runge2018causal,
  author  = {Runge, Jakob},
  title   = {Causal network reconstruction from time series: From theoretical assumptions to practical estimation},
  journal = {Chaos},
  volume  = {28},
  number  = {7},
  pages   = {075310},
  year    = {2018},
  doi     = {10.1063/1.5025050}
}

@article{nicoletti2026balancing,
  title={Balancing information and dissipation with partially observed fluctuating signals},
  author={Nicoletti, Giorgio and Di Terlizzi, Ivan and Busiello, Daniel Maria},
  journal={Physical Review Letters},
  volume={137},
  number={12},
  pages={127101},
  year={2026},
  publisher={APS}
}

@article{aoki2019universal,
  title={A universal biomolecular integral feedback controller for robust perfect adaptation},
  author={Aoki, Stephanie K and Lillacci, Gabriele and Gupta, Ankit and Baumschlager, Armin and Schweingruber, David and Khammash, Mustafa},
  journal={Nature},
  volume={570},
  number={7762},
  pages={533--537},
  year={2019},
  publisher={Nature Publishing Group UK London}
}

@article{pridham2024dynamical,
  title={Dynamical network stability analysis of multiple biological ages provides a framework for understanding the aging process},
  author={Pridham, Glen and Rutenberg, Andrew D},
  journal={The Journals of Gerontology, Series A: Biological Sciences and Medical Sciences},
  volume={79},
  number={10},
  pages={glae021},
  year={2024},
  publisher={Oxford University Press US}
}

@article{schaffer2021mapping,
  title={Mapping the multiscale structure of biological systems},
  author={Schaffer, Leah V and Ideker, Trey},
  journal={Cell systems},
  volume={12},
  number={6},
  pages={622--635},
  year={2021},
  publisher={Elsevier}
}

@article{de2023more,
  title={More is different in real-world multilayer networks},
  author={De Domenico, Manlio},
  journal={Nature Physics},
  volume={19},
  number={9},
  pages={1247--1262},
  year={2023},
  publisher={Nature Publishing Group UK London}
}

@article{nicoletti2024information,
  title={Information propagation in multilayer systems with higher-order interactions across timescales},
  author={Nicoletti, Giorgio and Busiello, Daniel Maria},
  journal={Physical Review X},
  volume={14},
  number={2},
  pages={021007},
  year={2024},
  publisher={APS}
}

@article{tostevin2009mutual,
  title={Mutual information between input and output trajectories of biochemical networks},
  author={Tostevin, Filipe and Ten Wolde, Pieter Rein},
  journal={Physical review letters},
  volume={102},
  number={21},
  pages={218101},
  year={2009},
  publisher={APS}
}

@footnote{supplemental_material,
  key={supplemental_material}, 
  note={See supplemental material for mathematical details, analytical derivations, and model details.}
}

@article{lan2012energy,
  title={The energy--speed--accuracy trade-off in sensory adaptation},
  author={Lan, Ganhui and Sartori, Pablo and Neumann, Silke and Sourjik, Victor and Tu, Yuhai},
  journal={Nature physics},
  volume={8},
  number={5},
  pages={422--428},
  year={2012},
  publisher={Nature Publishing Group UK London}
}

@article{frishman2020learning,
  title={Learning force fields from stochastic trajectories},
  author={Frishman, Anna and Ronceray, Pierre},
  journal={Physical Review X},
  volume={10},
  number={2},
  pages={021009},
  year={2020},
  publisher={APS}
}

@article{bo2017multiple,
  title={Multiple-scale stochastic processes: decimation, averaging and beyond},
  author={Bo, Stefano and Celani, Antonio},
  journal={Physics reports},
  volume={670},
  pages={1--59},
  year={2017},
  publisher={Elsevier}
}

@article{di2023variance,
author = {I. Di Terlizzi  and M. Gironella  and D. Herraez-Aguilar  and T. Betz  and F. Monroy  and M. Baiesi  and F. Ritort },
title = {Variance sum rule for entropy production},
journal = {Science},
volume = {383},
number = {6686},
pages = {971-976},
year = {2024},
doi = {10.1126/science.adh1823},
URL = {https://www.science.org/doi/abs/10.1126/science.adh1823},
eprint = {https://www.science.org/doi/pdf/10.1126/science.adh1823}
}

@article{nicoletti2025optimal,
	article_type = {journal},
	author = {Nicoletti, Giorgio and Bruzzone, Matteo and Suweis, Samir and dal Maschio, Marco and Busiello, Daniel Maria},
	citation = {eLife 2025;13:RP99767},
	doi = {10.7554/eLife.99767},
	editor = {Murugan, Arvind and Walczak, Aleksandra M},
	issn = {2050-084X},
	journal = {eLife},
	month = {jul},
	pages = {RP99767},
	pub_date = {2025-07-28},
	publisher = {eLife Sciences Publications, Ltd},
	title = {Optimal information gain at the onset of habituation to repeated stimuli},
	url = {https://doi.org/10.7554/eLife.99767},
	volume = 13,
	year = 2025}

@article{zeraati2026neural,
  title={Neural timescales from a computational perspective},
  author={Zeraati, Roxana and Levina, Anna and Macke, Jakob H and Gao, Richard},
  journal={Nature Neuroscience},
  pages={1--14},
  year={2026},
  publisher={Nature Publishing Group US New York}
}

@article{eckert2024biochemically,
  title={Biochemically plausible models of habituation for single-cell learning},
  author={Eckert, Lina and Vidal-Saez, Maria Sol and Zhao, Ziyuan and Garcia-Ojalvo, Jordi and Martinez-Corral, Rosa and Gunawardena, Jeremy},
  journal={Current Biology},
  volume={34},
  number={24},
  pages={5646--5658},
  year={2024},
  publisher={Elsevier}
}

@article{gao2020neuronal,
  title={Neuronal timescales are functionally dynamic and shaped by cortical microarchitecture},
  author={Gao, Richard and Van den Brink, Ruud L and Pfeffer, Thomas and Voytek, Bradley},
  journal={elife},
  volume={9},
  pages={e61277},
  year={2020},
  publisher={eLife Sciences Publications, Ltd}
}

@article{golesorkhi2021brain,
  title={The brain and its time: intrinsic neural timescales are key for input processing},
  author={Golesorkhi, Mehrshad and Gomez-Pilar, Javier and Zilio, Federico and Berberian, Nareg and Wolff, Annemarie and Yagoub, Mustapha CE and Northoff, Georg},
  journal={Communications biology},
  volume={4},
  number={1},
  pages={970},
  year={2021},
  publisher={Nature Publishing Group UK London}
}

@article{o2023nonlinear,
  title={Nonlinear slow-timescale mechanisms in synaptic plasticity},
  author={O'Donnell, Cian},
  journal={Current opinion in neurobiology},
  volume={82},
  pages={102778},
  year={2023},
  publisher={Elsevier}
}

@article{brandman2008feedback,
  title={Feedback loops shape cellular signals in space and time},
  author={Brandman, Onn and Meyer, Tobias},
  journal={Science},
  volume={322},
  number={5900},
  pages={390--395},
  year={2008},
  publisher={American Association for the Advancement of Science}
}

@article{brandman2005interlinked,
  title={Interlinked fast and slow positive feedback loops drive reliable cell decisions},
  author={Brandman, Onn and Ferrell Jr, James E and Li, Rong and Meyer, Tobias},
  journal={Science},
  volume={310},
  number={5747},
  pages={496--498},
  year={2005},
  publisher={American Association for the Advancement of Science}
}

@article{nicoletti2024gaussian,
  title={Information propagation in Gaussian processes on multilayer networks},
  author={Nicoletti, Giorgio and Busiello, Daniel Maria},
  journal={Journal of Physics: Complexity},
  volume={5},
  number={4},
  pages={045004},
  year={2024},
  publisher={IOP Publishing}
}

@article{nicoletti2025stochastic,
  title={Stochastic processes with multiple temporal scales: timescale separation and information},
  author={Nicoletti, Giorgio and Busiello, Daniel Maria},
  journal={Journal of Statistical Mechanics: Theory and Experiment},
  volume={2025},
  number={12},
  pages={124004},
  year={2025},
  publisher={IOP Publishing}
}

@article{Dahlhaus2000,
  author  = {Dahlhaus, Rainer},
  title   = {Graphical interaction models for multivariate time series},
  journal = {Metrika},
  volume  = {51},
  number  = {2},
  pages   = {157--172},
  year    = {2000},
  doi     = {10.1007/s001840000055}
}

@book{Brillinger1981,
  author    = {Brillinger, David R.},
  title     = {Time Series: Data Analysis and Theory},
  edition   = {Expanded},
  publisher = {Holden-Day},
  address   = {San Francisco},
  year      = {1981}
}

@article{ren2010noise,
  author  = {Ren, Jie and Wang, Wen-Xu and Li, Baowen and Lai, Ying-Cheng},
  title   = {Noise Bridges Dynamical Correlation and Topology in Coupled Oscillator Networks},
  journal = {Physical Review Letters},
  volume  = {104},
  number  = {5},
  pages   = {058701},
  year    = {2010},
  doi     = {10.1103/PhysRevLett.104.058701}
}

@article{auconi2017causal,
  author  = {Auconi, Andrea and Giansanti, Andrea and Klipp, Edda},
  title   = {Causal influence in linear {L}angevin networks without feedback},
  journal = {Physical Review E},
  volume  = {95},
  number  = {4},
  pages   = {042315},
  year    = {2017},
  doi     = {10.1103/PhysRevE.95.042315}
}

@article{baldovin2020understanding,
  author  = {Baldovin, Marco and Cecconi, Fabio and Vulpiani, Angelo},
  title   = {Understanding causation via correlations and linear response theory},
  journal = {Physical Review Research},
  volume  = {2},
  number  = {4},
  pages   = {043436},
  year    = {2020},
  doi     = {10.1103/PhysRevResearch.2.043436}
}

@article{smirnov2013spurious,
  author  = {Smirnov, Dmitry A.},
  title   = {Spurious causalities with transfer entropy},
  journal = {Physical Review E},
  volume  = {87},
  number  = {4},
  pages   = {042917},
  year    = {2013},
  doi     = {10.1103/PhysRevE.87.042917}
}

@article{james2016information,
  author  = {James, Ryan G. and Barnett, Nix and Crutchfield, James P.},
  title   = {Information Flows? {A} Critique of Transfer Entropies},
  journal = {Physical Review Letters},
  volume  = {116},
  number  = {23},
  pages   = {238701},
  year    = {2016},
  doi     = {10.1103/PhysRevLett.116.238701}
}

@article{runge2012escaping,
  author  = {Runge, Jakob and Heitzig, Jobst and Petoukhov, Vladimir and Kurths, J{\"u}rgen},
  title   = {Escaping the Curse of Dimensionality in Estimating Multivariate Transfer Entropy},
  journal = {Physical Review Letters},
  volume  = {108},
  number  = {25},
  pages   = {258701},
  year    = {2012},
  doi     = {10.1103/PhysRevLett.108.258701}
}

@article{deronde2010effect,
  author  = {{de Ronde}, Wiet Hendrik and Tostevin, Filipe and {ten Wolde}, Pieter Rein},
  title   = {Effect of feedback on the fidelity of information transmission of time-varying signals},
  journal = {Physical Review E},
  volume  = {82},
  number  = {3},
  pages   = {031914},
  year    = {2010},
  doi     = {10.1103/PhysRevE.82.031914}
}

@article{deronde2012feedforward,
  author  = {{de Ronde}, Wiet Hendrik and Tostevin, Filipe and {ten Wolde}, Pieter Rein},
  title   = {Feed-forward loops and diamond motifs lead to tunable transmission of information in the frequency domain},
  journal = {Physical Review E},
  volume  = {86},
  number  = {2},
  pages   = {021913},
  year    = {2012},
  doi     = {10.1103/PhysRevE.86.021913}
}

@article{mattingly2021escherichia,
  author  = {Mattingly, Henry H. and Kamino, Keita and Machta, Benjamin B. and Emonet, Thierry},
  title   = {{E}scherichia coli chemotaxis is information limited},
  journal = {Nature Physics},
  volume  = {17},
  number  = {12},
  pages   = {1426--1431},
  year    = {2021},
  doi     = {10.1038/s41567-021-01380-3}
}

@article{horowitz2014thermodynamics,
  author  = {Horowitz, Jordan M. and Esposito, Massimiliano},
  title   = {Thermodynamics with Continuous Information Flow},
  journal = {Physical Review X},
  volume  = {4},
  number  = {3},
  pages   = {031015},
  year    = {2014},
  doi     = {10.1103/PhysRevX.4.031015}
}

@article{matsumoto2018role,
  author  = {Matsumoto, Takumi and Sagawa, Takahiro},
  title   = {Role of sufficient statistics in stochastic thermodynamics and its implication to sensory adaptation},
  journal = {Physical Review E},
  volume  = {97},
  number  = {4},
  pages   = {042103},
  year    = {2018},
  doi     = {10.1103/PhysRevE.97.042103}
}

\clearpage
\newpage

\onecolumngrid
\begin{center}
\vspace{0.5cm}
\textbf{\large End Matter}
\end{center}
\twocolumngrid

\subsection{MIRMO elements for connected dofs do not change with the structure of the interaction tree}
\noindent Consider two directly connected dofs in a tree-like interaction network, a sender $x_{\rm S}$ and a receiver $x_{\rm R}$ evolving on their respective timescales $\tau_{\rm S}$ and $\tau_{\rm R}$. The label associated with their rescaled mutual information rate, $\mathcal{L}_{\rm SR}$, is independent of the surrounding tree topology. In other words, in a tree-like structure, $\mathcal{L}_{\rm SR}$ is solely dictated by the interaction between $x_{\rm S}$ and $x_{\rm R}$ and their relative timescales. Specifically, taking $x_{\rm S} \Rightarrow x_{\rm R}$, if $\tau_{\rm S} \ll \tau_{\rm R}$, then we have a direct interaction and $\mathcal{L}_{\rm SR} = 1$. Conversely, if $\tau_{\rm S} \gg \tau_{\rm R}$, then we have a feedback interaction and $\hat R_{\rm SR} \to +\infty$ in the timescale separation limit.

To prove it, consider a tree-like interaction network of coupled processes. Given one of these dofs, $x_k$, let $\mathcal{P}(k)$ be the set of indices of its ancestors, i.e., nodes with a directed pathway to $k$, including $k$ itself. In a tree, every such pathway is unique. Rescaling the time by an arbitrary timescale $\tau_\mu$, with the reference index $\mu$ held fixed as $r$ varies, and recalling that $\tau_\mu/\tau_k = r^{\mu-k}$, the spectrum of $x_k$, in the units of rescaled frequencies $\omega' = \omega \tau_\mu$, reads:
\begin{equation}
    S_{kk}(\omega') = \sum_{i \in \mathcal{P}(k)} S^{(0)}_{i}(\omega') |T_{ki}|^2
\end{equation}
with $T_{kk}=1$, where $S_i^{(0)} = 2D_i r^{\mu-i}/(\omega'^2 + r^{2(\mu-i)})$, and
\begin{equation}
    T_{ki} = \prod_{u\Rightarrow v \in \pi(i,k)} \frac{a_{vu} ~r^{\mu-v}}{r^{\mu-v} + i\omega'}
    \label{eq:T}
\end{equation}
where the product runs over the edges of the unique directed pathway $\pi(i,k)$ from $i$ to $k$. Since $|r^{\mu-v}/(r^{\mu-v} + i\omega')|\leq1$, we have:
\begin{equation}
    S_{kk}(\omega') \leq \sum_{i \in \mathcal{P}(k)} g^2_{ki} S^{(0)}_{i}(\omega')
    \label{eq:Skk}
\end{equation}
where $g_{ki}$ is the product of all couplings along the pathway from $i$ to $k$, with $g_{kk} = 1$. Integrating over the rescaled frequencies, we thus have:
\begin{equation}
    \langle x_k^2 \rangle = \frac{1}{2 \pi} \int_{-\infty}^{+\infty} S_{kk}(\omega') d\omega' \leq \sum_{i \in \mathcal{P}(k)} g^2_{ki} D_i = v_k \;.
    \label{eq:average}
\end{equation}
\textbf{Direct link}. Consider now a link $x_{\rm S} \Rightarrow x_{\rm R}$ with $\tau_R$ with $\tau_{\rm R} > \tau_{\rm S}$. If both $x_{\rm S}$ and $x_n$ drive $x_{\rm R}$, they cannot share an ancestor, since this would create a cycle in the underlying undirected network. Therefore, different inputs on the same dof are uncorrelated. Rescaling both time and frequency by $\tau_{\rm R}$, we have:
\begin{equation}
    \label{eq:phi}
    \phi_{\rm SR} = \frac{|S_{\rm SR}|^2}{S_{\rm SS}S_{\rm RR}} = \frac{a_{\rm RS}^2 S_{\rm SS}}{a_{\rm RS}^2 S_{\rm SS} + \sum_{k\neq {\rm S}} a_{{\rm R}k}^2 S_{kk} + 2 D_{\rm R}}
\end{equation}\\
where $|S_{\rm SR}|^2 = S_{\rm SS}^2 a^2_{\rm RS} /(1+\omega'^2)$ and $k$ runs over the drivers of ${\rm R}$. From Eq.~\eqref{eq:phi}, we readily obtain that $1-\phi_{\rm SR} \geq 2D_{\rm R}/(a_{\rm RS}^2 S_{\rm SS} + 2 D_R)$, so that $-\log(1-\phi_{\rm SR}) \leq a_{\rm RS}^2 S_{\rm SS}/2 D_R$. Integrating over the rescaled frequencies, and using \eqref{eq:average}, we obtain an upper bound on the rescaled mutual information rate between $x_{\rm S}$ and $x_{\rm R}$:
\begin{equation}
    \hat R_{\rm SR} \leq \frac{a_{\rm RS}^2}{4 D_{\rm R}} \left( \sum_{k \in \mathcal{P}({\rm S})} g^2_{{\rm S}k} D_{k} \right) = \frac{a_{\rm RS}^2}{4 D_{\rm R}} \,v_{\rm S}\;.
    \label{eq:upperbound}
\end{equation}
Analogously, we can obtain a lower bound by noting that:
\begin{equation}
    \hat{R}_{\rm SR} \geq - \frac{1}{2 \pi} \int_{1}^{+\infty} \log(1-\phi_{\rm SR}) d\omega' \geq \frac{1}{2 \pi} \int_{1}^{+\infty} \phi_{\rm SR} d\omega' \;.
    \label{eq:lowerR}
\end{equation}
Since, from Eq.~\eqref{eq:Skk}, when $\omega' \geq 1$, $S_{kk} \leq 2 v_k/r$ for all $k\neq R$, this inequality sets an upper bound on the denominator of the coherence in Eq.~\eqref{eq:phi}. Further, we can lower bound its numerator by noting that $S_{\rm SS} \geq S^{(0)}_{\rm S}$. Therefore, inserting this back into Eq.~\eqref{eq:lowerR}, solving the integral over rescaled frequencies, and taking the timescale separation limit, $r\to +\infty$, we obtain:
\begin{equation}
    \hat{R}_{\rm SR} \geq \frac{a_{\rm RS}^2}{4 D_R} \, D_{\rm S} \;.
    \label{eq:lowerbound}
\end{equation}
Equations \eqref{eq:upperbound} and \eqref{eq:lowerbound} together determine that the rescaled mutual information rate stays bounded above and away from zero in the presence of direct link, independently of the surrounding interaction tree, thereby $\mathcal{L}_{SR} = 1$.

\noindent\textbf{Feedback link}. A similar derivation can be performed in the presence of a feedback link between $x_{\rm S}$ and $x_{\rm R}$. This case is characterized by a connection $x_{\rm S} \Rightarrow x_{\rm R}$ with $\tau_{\rm S} > \tau_{\rm R}$. In this case, we need to rescale both time and frequency by $\tau_{\rm S}$ so that the coherence retains a term in the denominator that depends on $\tau_{\rm R}/\tau_{\rm S}$. In fact, we have:
\begin{equation}
    \label{eq:phiFEED}
    \phi_{\rm SR} = \frac{a_{\rm RS}^2 S_{\rm SS}}{a_{\rm RS}^2 S_{\rm SS} + \sum_{k\neq {\rm S}} a_{{\rm R}k}^2 S_{kk} + 2 D_{\rm R}\tau_{\rm R}/\tau_{\rm S}} \;.
\end{equation}
To derive a lower bound for the $\hat{R}_{\rm SR}$, we write:
\begin{equation}
    (1 - \phi_{\rm SR})^{-1} \geq 1 + \frac{r}{2} \frac{a_{\rm RS}^2 S_{\rm SS}}{D_{R} + \sum_{k\neq {\rm S}} a_{{\rm R}k}^2 v_{k}}
\end{equation}
by using that $S_{kk} \leq 2 v_k/r$ for $\omega' \geq 1$ and $k\neq S$, and that $\tau_{\rm R}/\tau_{\rm S} \leq 1/r$. In the same frequency range, $S_{\rm SS} \geq S^{(0)}_S = 2 D_{\rm S}/(1+\omega'^2)\geq D_S/\omega'^2$, so that:
\begin{equation}
    \hat{R}_{\rm SR} \geq \frac{1}{2 \pi}\int_1^{+\infty} \log\left(1 + \frac{r}{2} \frac{a_{\rm RS}^2 D_{\rm S} (\omega')^{-2}}{D_{R} + \sum_{k\neq {\rm S}} a_{{\rm R}k}^2 v_{k}}\right)d\omega'\;.
\end{equation}
Since all terms are positive, direct computation gives:
\begin{equation}
    \hat{R}_{\rm SR} \xrightarrow{r\to\infty} +\infty
\end{equation}
since its lower bound diverges for $r \to +\infty$.

\subsection{MIRMO and the conditional mutual information rate uniquely reconstruct the interaction tree}
\noindent We are now ready to show that, by combining MIRMO and the conditional mutual information rate, we are always able to determine exactly the underlying interaction tree governing a set of multiscale processes.

\noindent\textbf{Mutual information rate.} Consider a tree of interacting dofs. In general terms, $\hat{R}_{kl}$ is zero if and only if $S_{kl}(\omega')$ vanishes identically. For a tree-like structure, we have:
\begin{equation}
    S_{kl}(\omega') = \sum_{i\in \mathcal{P}(k)\cap \mathcal{P}(l)} T_{ki}(\omega') T_{li}^*(\omega') S^{(0)}_i(\omega')
\end{equation}
If $\mathcal{P}(k)\cap\mathcal{P}(l)$ is empty, this sum vanishes. Otherwise, let $b$ be a common ancestor at a branching point of the pathway between $k$ and $l$, including the cases of direct connection, $b = k$ or $b = l$. The total shared ancestors are exactly $\mathcal{P}(b)$. Therefore, $S_{kl} = T_{kb}T_{lb}^*S_{bb} \neq 0$ and $\hat{R}_{kl}$ is nonzero. Therefore, the rescaled mutual information rate is positive, at finite $r$, between two dofs that are directly connected or share a common ancestor.

\noindent\textbf{Conditional mutual information rate.} In a tree of interacting dofs, $R_{kl|-kl}$ is greater than zero at finite $r$ if and only if $x_k$ and $x_l$ are linked or if they both drive a common dof. As for $\hat{R}_{kl}$, these properties can be derived from the general expression of $P_{kl}$ starting from the definition:
\begin{equation}
    P = S^{-1} = C^\dagger S_0^{-1} C \;,
\end{equation}
where the $kl$-th element of $S$ is $S_{kl}$, $(S_0)_{kk} = S^{(0)}_k$, and the $kl$-th of $C$ is $C_{kl} = \delta_{kl} - a_{kl} h_k$ with $h_k = r^{\mu-k}/(r^{\mu-k} + i \omega')$, where $\mu$ is the reference node to rescale the time, as above. Specifying this expression to the case of $k \neq l$, we obtain:
\begin{equation}
    P_{kl} = -\frac{a_{kl} h_k}{S_k^{(0)}} -\frac{a_{lk} h^*_l}{S_l^{(0)}} +\sum_{s\neq k,l} \frac{a_{sk} a_{sl} |h_s|^2}{S^{(0)}_s} \;.
    \label{eq:Pkl}
\end{equation}
$R_{kl|-kl}$ is zero only when $P_{kl}$ vanishes identically. Eq.~\eqref{eq:Pkl} contains terms only if $x_k$ and $x_l$ are directly connected or if they appear in the same equation of motion. The tree structure is crucial to avoid cancellations in the summation of Eq.~\eqref{eq:Pkl}: if $x_k$ and $x_l$ are connected, they cannot drive a common node $x_s$, while if they drive a common node, that node is unique.

\begin{table}[t!]
    \centering
    \begin{tabular}{l|c|c}
        \textit{Conditions} & \, $\hat{R}_{kl}$ \, & \, $\hat{R}_{kl|-kl}$ \, \\
        \hline
        $x_k$ and $x_l$ are linked & $>0$ & $>0$ \\
        $x_k$ and $x_l$ drive a common node & $0$ & $>0$\\
        $x_k$ and $x_l$ have a common ancestor & $>0$ & $0$\\
        None of the above & $0$ & $0$
    \end{tabular}
    \caption{Behavior of the mutual information rate and the conditional mutual information rate with different interaction topologies at an arbitrary finite $r$. Positive entries need not have positive limits as $r\to+\infty$.}
    \label{tab:conditions}
\end{table}

\noindent\textbf{Bijection.} We can summarize the rules derived above in Table \ref{tab:conditions}. It follows that all interaction links are those for which both $R_{kl}$ and $R_{kl|-kl}$ are positive. This identification is independent of the $r\to +\infty$ limit, but it does not contain any information on the directionality. When replacing the mutual information rate with the MIRMO, interaction links correspond to all cases in which the labels are nonzero. Since the tree structure does not affect these labels, $\mathcal{L}_{kl}$ determines the directions of the reconstructed links in the timescale separation limit. The fact that dofs driving a common node have only a positive conditional mutual information rate rules out any possible ambiguities. Therefore, to establish the bijection, it is sufficient to evaluate the rescaled mutual information rate only through its labels in the timescale separation limit.

\noindent\textbf{Fork-free trees.} Topologically, each case in which one node drives more than one dof constitutes a fork structure. Therefore, restricting the analysis to fork-free interaction trees amounts to considering tree-like networks in which each node drives at most one other dof. In this simplified scenario, when we ignore the slowest node $x_M$ of a given piece while retaining the original pairwise labels, nodes belonging to different remaining pieces have label $0$. Either they share no information, or their information passes through $x_M$ along a path containing a direct link into it. In the latter case, their pairwise rescaled rate vanishes in the timescale separation limit, since the unique directed path passes through $x_M$, which is slower than both endpoints. Meanwhile, every actual edge inside a piece retains a positive label. Therefore, taking connected components of the positive-label graph identifies exactly the connected pieces left after removing $x_M$. We repeat the splitting until only single nodes or pairs of nodes remain, so that the connection between positive labels and actual links is unambiguous. Then, we rebuild the topology in reverse order. If a remaining sub-network feeds into the restored node, its pairwise labels with that node are $0$ or $1$, with at least one $1$, and it has to be attached through the unique node where its internal arrows end. If the restored node feeds into a remaining sub-network, its pairwise labels are $0$ or $2$, with at least one $2$. The nodes labeled $2$ form the chain it influences, and the restored node has to be attached to the beginning of that labeled chain in the already reconstructed sub-network. There can be at most one such outgoing attachment. Every reconnection is therefore uniquely determined. Since we used only labels, MIRMO alone is sufficient to reconstruct exactly the interaction tree within the fork-free class.

\clearpage
\onecolumngrid
\begin{center}
  {\large\bfseries Supplemental Material for ``Resolving the backbone of multiscale interactions through mutual information rate''}\\[2.5ex]
  Alberto Raiola,$^{1,2}$ Giorgio Nicoletti,$^{3,4,*}$ and Daniel M. Busiello$^{1,5,4,*}$\\[1.5ex]
  {\small
  $^1$Department of Physics and Astronomy ``G.\ Galilei'', University of Padova, 35131 Padova, Italy\\
  $^2$INFN, Padova Section, 35131 Padova, Italy\\
  $^3$Quantitative Life Sciences section, The Abdus Salam International Center for Theoretical Physics (ICTP), Trieste, Italy\\
  $^4$Theoretical Science Visiting Program (TSVP), Okinawa Institute of Science and Technology, Onna, Okinawa, Japan\\
  $^5$Max Planck Institute for the Physics of Complex Systems, 01187 Dresden, Germany\\[1ex]
  $^*$D.M.B.\ and G.N.\ contributed equally to this work}
\end{center}

\setcounter{equation}{0}\setcounter{figure}{0}\setcounter{table}{0}\setcounter{section}{0}
\renewcommand{\theequation}{S\arabic{equation}}
\renewcommand{\thefigure}{S\arabic{figure}}
\renewcommand{\thetable}{S\arabic{table}}

\section{Notation and definitions}
\label{sec:notation}
\noindent In this section we introduce the notation used throughout this work. Let $\mathcal{X} = \{x_1,x_2,\dots,x_N\}$ be the set of $N$ degrees of freedom (dofs) ruled by the following Langevin equations:
\begin{align}
\tau_1 \frac{dx_1}{dt} &=
-x_1 + \sum_{i\neq 1} f_{1i}(x_i)
+ \sqrt{2D_1\tau_1}\,\xi_1(t)
\nonumber\\
\tau_2 \frac{dx_2}{dt} &=
-x_2 + \sum_{i\neq 2} f_{2i}(x_i)
+ \sqrt{2D_2\tau_2}\,\xi_2(t)
\nonumber\\
&\vdots
\nonumber\\
\tau_N \frac{dx_N}{dt} &=
-x_N + \sum_{i=1}^{N-1} f_{Ni}(x_i)
+ \sqrt{2D_N\tau_N}\,\xi_N(t).
\label{eq:langevin_system}
\end{align}
Here, \(D_1,\dots,D_N\) denote the diffusion coefficients while
\(\tau_1,\dots,\tau_N\) are the characteristic timescales of the processes.
The functions \(f_{ij}\) encode the couplings \(x_j \Rightarrow x_i\) and generally may be linear or nonlinear.
The stochastic terms \(\xi_i(t)\) are independent Gaussian white noises satisfying $ \langle \xi_i(t)\rangle = 0 $ and $\langle \xi_i(t)\xi_j(t')\rangle= \delta_{ij}\delta(t-t').$ The timescales are assumed to be hierarchically ordered,
\begin{equation}
    \tau_1 \ll \tau_2 \ll \dots \ll \tau_N.
\end{equation}
Throughout this work we consider the parametrization
\begin{equation}
    \tau_i = r^{i-1}\tau_1
\end{equation}
with $r>1$. The limit of infinite timescale separation is therefore defined as $r\to\infty$. For each $i$, we denote the whole trajectory of $x_i$ by $\bm{x}_{i,T} = \{x_i(t)\}_{0\le t\le T}$.

For analytical calculations, all coupling functions are assumed to be linear, namely
\begin{equation}
    f_{ij}(x_j) = a_{ij} x_j
\end{equation}
where $a_{ij}$ denotes the strength of the coupling $x_j \Rightarrow x_i$ and $a_{ii}=0$. For analytical calculations, we consider stationary dynamics with $D_i>0$ and fixed non-zero couplings on the existing links. The numerical simulations in the main text also consider nonlinear couplings of the form $f_{ij}(x_j) = a_{ij} f(x_j)$.

\subsection{Mutual information rate}
\noindent The main quantity we consider in this work is the mutual information rate between two dofs $x_i$ and $x_j$, as defined in the main text,
\begin{equation}
R_{ij} = \lim_{T\to\infty}\frac{1}{T} I(\bm{x}_{i,T};\bm{x}_{j,T}),
\label{eq:mir_def}
\end{equation}
which, for Gaussian processes, reads
\begin{equation}
R_{ij} = -\frac{1}{4\pi}\int_{-\infty}^{+\infty} d\omega \, \log\left(1-\phi_{ij}(\omega)\right), \qquad \phi_{ij}(\omega) = \frac{|S_{ij}(\omega)|^2}{S_{ii}(\omega)S_{jj}(\omega)},
\label{eq:mir_gauss}
\end{equation}
where $\phi_{ij}$ is the coherence and $S_{ij}(\omega)$ is the $ij$-th element of the power spectrum.
Since \(R_{ij}\) has units of inverse time, it is convenient to introduce a dimensionless quantity before taking the limit \(r\to\infty\). Thus, we rescale the time with respect to the timescale \(\tau_\mu\) of a reference dof \(x_\mu\), with \(1 \le \mu \le N\). We use Greek indices for the reference choice, which is held fixed as $r$ varies, and Latin indices for generic dofs. We therefore introduce the dimensionless time variable $t' = \frac{t}{\tau_{\mu}}$. Under this transformation,
\begin{equation}
    \frac{dx_i}{dt}
=
\frac{1}{\tau_{\mu}}\frac{dx_i}{dt'}
=
\frac{1}{\tau_{\mu}}\dot{x}_i,
\end{equation}
with the noise terms rescaled accordingly,
\begin{equation}
    \xi_i(t)
=
\frac{1}{\sqrt{\tau_{\mu}}}\,
\xi_i(t')
\end{equation}
so that the rescaled noises satisfy $\langle \xi_i(t_1')\xi_j(t_2')\rangle= \delta_{ij}\delta(t_1'-t_2')$. Substituting these transformations into Eq.~\eqref{eq:langevin_system}, one obtains:
\begin{align}
\dot{x}_1 &=
-r^{\mu-1}x_1
+
\sum_{i\neq 1}
r^{\mu-1}f_{1i}(x_i)
+
\sqrt{2D_1r^{\mu-1}}\,\xi_1(t')
\nonumber\\
\dot{x}_2 &=
-r^{\mu-2}x_2
+
\sum_{i\neq 2}
r^{\mu-2}f_{2i}(x_i)
+
\sqrt{2D_2r^{\mu-2}}\,\xi_2(t')
\nonumber\\
&\vdots
\nonumber\\
\dot{x}_{\mu} &=
-x_{\mu}
+
\sum_{i\neq \mu}f_{\mu i}(x_i)
+
\sqrt{2D_{\mu}}\,\xi_{\mu}(t')
\nonumber\\
&\vdots
\nonumber\\
\dot{x}_N &=
-r^{\mu-N}x_N
+
\sum_{i\neq N}
r^{\mu-N}f_{Ni}(x_i)
+
\sqrt{2D_Nr^{\mu-N}}\,\xi_N(t').
\label{eq:rescaled_system}
\end{align}
The corresponding dimensionless mutual information rate, measured in units of the reference timescale \(\tau_{\mu}\), is then
\begin{equation}
\hat R^{(\mu)}_{ij}
=
\tau_{\mu} R_{ij}
=
-\frac{1}{4\pi}
\int_{-\infty}^{+\infty}
\log(1-\phi_{ij}(\omega'))\,d\omega',
\label{eq:rescaled_mir}
\end{equation}
where $\omega' = \tau_{\mu} \omega$ is the dimensionless frequency and, with a slight abuse of notation, $\phi_{ij}(\omega')$ denotes the coherence expressed as a function of $\omega'$. This newly defined quantity \(\hat R^{(\mu)}_{ij}\) admits a well-defined scaling with $r$ in the timescale-separation limit $r \to \infty$: it can vanish, converge to a finite value or diverge. The rescaled mutual information rate $\hat R_{ij}$ of the main text is obtained by rescaling with respect to the slower of the two timescales $\tau_i$ and $\tau_j$, i.e.,
\begin{equation}
\hat R_{ij} = \hat R^{(\mu)}_{ij},\qquad \mu=\max\{i,j\}.
\label{eq:hatR_main}
\end{equation}
Throughout Section~\ref{sec:elementary}, the other dofs are ordered as $k<l<m<\mu$ unless otherwise stated.

The spectra in the rescaled time obey
\begin{equation}
S^{(\mu)}_{ij}(\omega')=\frac{1}{\tau_\mu}S_{ij}\!\left(\frac{\omega'}{\tau_\mu}\right),
\qquad
P^{(\mu)}_{ij}(\omega')=\tau_\mu P_{ij}\!\left(\frac{\omega'}{\tau_\mu}\right).
\label{eq:spectral_rescaling}
\end{equation}
Here $P=S^{-1}$. We suppress the reference superscript on spectra written as functions of $\omega'$ below.

For linear couplings, we denote by $H_{ji}(\omega')$ the transfer function associated with a link $x_i \Rightarrow x_j$, i.e., the linear response of $\tilde x_j(\omega')$ to $\tilde x_i(\omega')$ in Fourier space. Note that, as for the coupling strength $a_{ji}$, the index of the receiving dof comes first. In physical-frequency units, the Fourier transform of Eq.~\eqref{eq:langevin_system} reads $\sum_jB_{ij}(\omega)\tilde x_j(\omega)=\tilde\eta_i(\omega)$, with
\begin{equation}
    B_{ij}(\omega)=(1+i\omega\tau_i)\,\delta_{ij}-a_{ij}
\end{equation}
where $\tilde\eta_i$ is the Fourier transform of the noise term $\sqrt{2D_i\tau_i}\,\xi_i$. Since the noises are independent, their spectra form the diagonal matrix $Q=\mathrm{diag}(q_1,\dots,q_N)$, with $q_i=2D_i\tau_i$. The power spectrum and its inverse are then given by
\begin{equation}
S(\omega)=B(\omega)^{-1}\,Q\,B(\omega)^{-\dagger},
\qquad
P(\omega)= S(\omega)^{-1}=B(\omega)^{\dagger}\,Q^{-1}\,B(\omega).
\label{eq:inverse_spectrum}
\end{equation}
We note that, for linear interactions, Eq.~\eqref{eq:inverse_spectrum} does not require timescale separation.

\subsection{Conditional mutual information rate}
\noindent The coherence between $x_i$ and $x_j$ obtained after conditioning on the trajectories of all the other dofs is known as \emph{partial coherence}, and reads
\begin{equation}
    \phi_{ij|-ij}(\omega)=\frac{|P_{ij}(\omega)|^2}{P_{ii}(\omega)P_{jj}(\omega)}
\end{equation}
where $-ij$ denotes all the dofs except $i$ and $j$ \cite{Dahlhaus2000, Brillinger1981}. In analogy with Eq.~\eqref{eq:mir_gauss}, the mutual information rate between $x_i$ and $x_j$, conditioned on the trajectories of all the other dofs, reads
\begin{equation}
R_{ij|-ij}=-\frac{1}{4\pi}\int_{-\infty}^{+\infty}d\omega\,\log\left(1-\phi_{ij|-ij}(\omega)\right).
\label{eq:conditional_mir}
\end{equation}
Like $R_{ij}$, the conditional mutual information rate is non-negative and it vanishes if and only if $P_{ij}(\omega)=0$ for almost all $\omega$.

\section{Elementary links}
\label{sec:elementary}
\noindent  Because interactions between dofs are assumed to be pairwise, the system can be naturally represented as a directed graph in which nodes correspond to variables and edges represent couplings. We define a \emph{direct link} as an interaction from node \(x_i\) to node \(x_j\) with \(i<j\), corresponding to a coupling from a faster variable to a slower one. Conversely, a \emph{feedback link} is an interaction from node \(x_i\) to node \(x_j\) with \(i>j\), corresponding to a coupling from a slower variable to a faster one.
Along a directed path, a direct link $x_i \Rightarrow x_j$ is said to be \emph{enclosed} if, before reaching this link, the path has already visited a dof slower than $x_j$. For instance, in $x_3 \Rightarrow x_1 \Rightarrow x_2$ the direct link $x_1 \Rightarrow x_2$ is enclosed, whereas in $x_2 \Rightarrow x_1 \Rightarrow x_3$ the direct link $x_1 \Rightarrow x_3$ is not.

In this section, we analyze the elementary building blocks of the interaction network, namely direct and feedback links, and determine how each of them contributes to the mutual information rate between pairs of variables. In particular, we focus on the asymptotic behavior of the rescaled mutual information rate $\hat R^{(\mu)}_{ij}$ as $r \to \infty$.

To describe the frequency range over which two dofs share fluctuations, we introduce the peak height $\mathcal A_{ij}(r)$ and the width $\Delta_{ij}(r)$ of their coherence. The elementary motifs considered below have their maximum at zero frequency, so we define
\begin{equation}
\mathcal A_{ij}(r)=\max_{\omega'}\phi_{ij}(\omega')=\phi_{ij}(0),\qquad
\phi_{ij}\!\left(\Delta_{ij}(r)\right)=\frac{\mathcal A_{ij}(r)}{2}.
\label{eq:coherence_peak_width}
\end{equation}
Here $\Delta_{ij}$ is the first positive frequency at which the coherence falls to half its peak value, i.e., the half-width at half-maximum in the dimensionless frequency $\omega'$. For a Lorentzian coherence, these definitions give $\phi_{ij}(\omega')=\mathcal A_{ij}/[1+(\omega'/\Delta_{ij})^2]$. When the coherence contains contributions on different frequency scales, we specify the peak and width of each contribution separately, since a broad weak component can contribute substantially to the integrated rate.

\subsection{Direct link ending in $\mu$}
\noindent Consider a direct link $x_m \Rightarrow x_{\mu}$, with $m < \mu$, so that
\begin{align}
    \dot x_m &= -r^{\mu-m}x_m + \sqrt{2D_mr^{\mu-m}}\xi_m(t') \nonumber \\
    \dot x_{\mu} &= -x_{\mu}  + a_{\mu m}x_m + \sqrt {2D_{\mu}}\xi_{\mu}(t')
\end{align}
where $a_{\mu m} = a$ is the coupling constant (in this case, $f_{\mu m}(x_m) = a_{\mu m}x_m = ax_m$). Taking the Fourier transform of the previous equations gives
\begin{align}
     \tilde x_m(\omega') &= \frac{\sqrt{2D_mr^{\mu-m}}}{i\omega' + r^{\mu-m}}\Xi_m \nonumber \\
    \tilde  x_{\mu}(\omega') &= \frac{a}{i\omega' + 1} \tilde x_m(\omega') + \frac{\sqrt{2D_{\mu}}}{i\omega' +1}\Xi_{\mu}
\end{align}
where $\Xi_i$ is the Fourier transform of $\xi_i(t')$. Taking into account that $\langle \Xi_m\Xi_{\mu}^* \rangle = 0$, the power spectrum of the system is
\begin{align}
     S_{mm}(\omega') &=  \langle |\tilde x_m(\omega')|^2\rangle = \frac{2D_mr^{\mu-m}}{\omega'^2 + r^{2(\mu-m)}} = S^{(0)}_m(\omega') \nonumber \\
     S_{\mu\mu}(\omega') &=  \langle |\tilde x_{\mu}(\omega')|^2\rangle = \frac{a^2}{\omega'^2 + 1}\frac{2D_m r^{\mu-m}}{\omega'^2 + r^{2(\mu-m)}} + \frac{2D_{\mu}}{\omega'^2 + 1} = |H_{\mu m}(\omega')|^2 S^{(0)}_m(\omega') + S^{(0)}_{\mu}(\omega') \\
     S_{m\mu}(\omega') &= \langle \tilde x_m(\omega') \tilde x_{\mu}^*(\omega')\rangle = \frac{a}{1 - i\omega'}\frac{2D_m r^{\mu-m}}{\omega'^2 + r^{2(\mu-m)}} = H_{\mu m}^*(\omega') S^{(0)}_m(\omega') \nonumber
\end{align}
where we call $S^{(0)}_i(\omega')$ the natural spectrum of $x_i$, namely the power spectrum that $x_i$ would have in the absence of couplings:
\begin{equation}
    S^{(0)}_i(\omega') = \frac{2D_ir^{\mu-i}}{\omega'^2 + r^{2(\mu-i)}}
\end{equation}
and $H_{\mu m}(\omega')$ is the transfer function of the link $x_m \Rightarrow x_{\mu}$. The coherence between the two signals is thus
\begin{equation}
    \phi_{m\mu}(\omega') = \frac{|S_{m\mu}(\omega')|^2}{S_{mm}(\omega')S_{\mu\mu}(\omega')} = |H_{\mu m}(\omega')|^2\frac{S_{mm}(\omega')}{S_{\mu\mu}(\omega')} = \frac{r^{\mu-m}}{r^{\mu-m} + \frac{D_{\mu}}{a^2 D_m} (r^{2(\mu-m)} + \omega'^2)}
\end{equation}
and the dimensionless mutual information rate is
\begin{equation}
    \hat R_{m\mu} = -\frac{1}{4 \pi}\int_{-\infty}^{+\infty} d \omega' \log(1- \phi_{m\mu}(\omega')) =  \frac{r^{\mu-m}}{2} \left (\sqrt{1 + \frac{a^2D_m}{D_{\mu}r^{\mu-m}}} - 1\right )  \underbrace{\to}_{r \to \infty } \frac{1}{4}\frac{a^2D_m}{D_{\mu}}
\end{equation}
that, in the limit of timescale separation, converges to a finite value independent of the timescale ratio, and which instead depends only on the coupling strength and on the ratio between the diffusion coefficients. This follows from the shape of the coherence function. In the timescale-separation limit $r \to \infty$, the whole coherence becomes small. Therefore, the logarithm can be expanded as $-\log(1 - \phi_{m\mu}(\omega')) \approx \phi_{m\mu}(\omega')$ and $\hat R_{m\mu}$ reduces to the integral of a Lorentzian function whose peak value scales with $r$ as $\mathcal{A}_{m\mu}(r) = \phi_{m\mu}(0) \sim 1/r^{\mu-m}$ and whose width scales as $\Delta_{m\mu}(r) \sim r^{\mu-m}$, i.e,
\begin{equation}
    \phi_{m\mu}(\omega') = \mathcal{A}_{m\mu}(r) g \left ( \frac{\omega'}{\Delta_{m\mu}(r)} \right )
\end{equation}
where $g(u) = 1/(1 + u^2)$ is a standard Lorentzian independent of $r$. Performing the change of variable $u = \omega'/\Delta_{m\mu}(r)$, we can write the following expression for the rate:
\begin{equation}
    \hat R_{m\mu} \approx \frac{1}{4\pi} \mathcal{A}_{m\mu}(r)\Delta_{m\mu}(r) \underbrace{\int_{-\infty}^{+\infty} g(u) du }_{\rm const} \sim \mathcal{A}_{m\mu}(r)\Delta_{m\mu}(r) \sim 1 
\end{equation}
which shows that $\hat R_{m\mu}$ approaches a finite value as $r \to \infty$.

By inspecting the graphs of $S_{mm}$ and $S_{\mu\mu}$, it is possible to appreciate the low-pass filtering action of a direct link. In particular, since the transfer function is
\begin{equation}
    |H_{\mu m}(\omega')|^2 = \frac{a^2}{\omega'^2 + 1}
\end{equation}
the frequencies $\omega' > 1$ coming from $x_m$ are attenuated and $S_{\mu\mu}(\omega')$ starts decreasing at $\omega' \approx 1$. For sufficiently large $r$, the transmitted contribution becomes negligible over most of the spectrum and we obtain
\begin{equation}
    S_{\mu\mu}(\omega') \approx \frac{2D_{\mu}}{\omega'^2 + 1} = S^{(0)}_{\mu}(\omega').
\end{equation}

\subsection{Direct link ending in $m<\mu$}
\label{direct_alto}
\noindent Consider now a direct link $x_l \Rightarrow x_m$  with $l < m < \mu$, while node $x_{\mu}$ is completely disconnected from the interacting pair. The equations of motion in this case are:
\begin{align}
    \dot x_l &= -r^{\mu-l}x_l + \sqrt{2D_lr^{\mu-l}}\xi_l(t') \nonumber\\
    \dot x_m &= -r^{\mu-m}x_m  + ar^{\mu-m}x_l + \sqrt {2D_mr^{\mu-m}}\xi_m(t')
\end{align}
and, proceeding as in the previous section, the power spectra are
\begin{align}
S_{ll}(\omega') &= \langle |\tilde x_l(\omega')|^2\rangle = \frac{2D_l r^{\mu-l}}{\omega'^2 + r^{2(\mu-l)}} \nonumber \\
S_{mm}(\omega') &= \langle |\tilde x_m(\omega')|^2\rangle = \frac{a^2r^{2(\mu-m)}}{\omega'^2 + r^{2(\mu-m)}}\frac{2D_l r^{\mu-l}}{\omega'^2 + r^{2(\mu-l)}} + \frac{2D_m r^{\mu-m}}{\omega'^2 + r^{2(\mu-m)}} \\
S_{ml}(\omega') &= \langle \tilde x_m(\omega')\tilde x_l^*(\omega')\rangle = \frac{a r^{\mu-m}}{i\omega' + r^{(\mu-m)}}\frac{2D_l r^{\mu-l}}{\omega'^2 + r^{2(\mu-l)}} \nonumber
\end{align}
with the corresponding coherence given by
\begin{equation}
\phi_{lm}(\omega') = \frac{r^{l+2\mu}}{r^{l+2\mu} + \frac{D_m}{a^2D_l}(r^{2\mu + m } + r^{2l + m}\omega'^2)} \; .
\end{equation}
Here, the rate rescaled with respect to $\tau_{\mu}$ diverges. Indeed, the peak height and the width of the coherence scale as  $\mathcal{A}_{lm}(r) \sim r^{-(m-l)}$ and $\Delta_{lm}(r) \sim r^{\mu-l}$ so that  $\hat R^{(\mu)}_{lm} \sim r^{\mu-m} \to \infty$. Since both $x_l$ and $x_m$ evolve on timescales much shorter than $x_{\mu}$, measuring the information rate in units of $\tau_{\mu}$ introduces a factor $\tau_{\mu}/\tau_m = r^{\mu-m}$ with respect to the finite rate measured in units of $\tau_m$, and this factor diverges as $r \to \infty$. As in the previous case, the direct link can be interpreted as a low-pass filter whose transfer function is
\begin{equation}
H_{ml}(\omega') = \frac{ar^{\mu-m}}{i\omega' + r^{(\mu-m)}}.
\end{equation}
The cutoff frequency is now of order $r^{\mu-m}$, reflecting the fact that the receiving variable evolves on a much faster timescale than $x_{\mu}$.

\subsection{Feedback link starting from $\mu$}
\noindent We now consider the elementary motif consisting of a single feedback link from the slowest variable $x_{\mu}$ to a faster variable $x_m$, $x_{\mu} \Rightarrow x_m$, corresponding to
\begin{align}
    \dot x_m &= -r^{\mu-m}x_m + a r^{\mu-m}x_{\mu}+ \sqrt{2D_mr^{\mu-m}}\xi_m(t') \nonumber\\
    \dot x_{\mu} &= -x_{\mu}   + \sqrt {2D_{\mu}}\xi_{\mu}(t')
\end{align}
and power spectra
\begin{align}
    S_{mm}(\omega') &=\langle |\tilde x_m(\omega')|^2 \rangle  = \frac{a^2 r^{2(\mu-m)}}{\omega'^2 + r^{2(\mu-m)}} \frac{2D_{\mu}}{\omega'^2 + 1} + \frac{2D_m r^{\mu-m}}{\omega'^2 + r^{2(\mu-m)}}  \nonumber
    \\
    S_{\mu\mu}(\omega') &= \langle |\tilde x_{\mu}(\omega')|^2 \rangle  = \frac{2D_{\mu}}{\omega'^2 + 1} \\
    S_{m\mu}(\omega') &=\langle \tilde x_m(\omega')\tilde x_{\mu}^*(\omega') \rangle  = \frac{ar^{\mu-m}}{i\omega' + r^{\mu-m}}\frac{2D_{\mu}}{\omega'^2 + 1} \nonumber
\end{align}
 with coherence
\begin{equation}
\phi_{m\mu}(\omega') =  \frac{r^{\mu-m}}{r^{\mu-m}+\frac{D_m}{a^2D_{\mu}}(\omega'^2+1)} \; .
\end{equation}
For a feedback link, the coherence approaches unity over an increasingly broad frequency range as $r \to \infty$. Consequently, the approximation $-\log(1 - \phi) \approx \phi$ is no longer valid. In this case the coherence has peak height $\mathcal{A}_{m\mu} \sim 1$ and width $\Delta_{m\mu} \sim r^{(\mu-m)/2}$. We can infer the order of divergence by performing the direct calculation. In particular, we use the relation
\begin{equation}
\int_{-\infty}^{+\infty} \log(x^2 + \alpha^2)dx = 2\pi \alpha + C
\end{equation}
where $\alpha \in \mathbb R^+$ and $C$ is a divergent constant independent of $\alpha$, understood with the same frequency cutoff in both integrals below, so that we obtain
\begin{align}
    \hat R_{m\mu} &= -\frac{1}{4 \pi} \int_{-\infty}^{+\infty}\log \left (\frac{\frac{D_m}{a^2D_{\mu}} (\omega'^2 + 1)}{r^{\mu-m} + \frac{D_m}{a^2D_{\mu}} (\omega'^2 + 1)}  \right ) d\omega' = \nonumber\\
    &= -\frac{1}{4\pi} \int_{-\infty}^{+\infty} \log(\frac{D_m}{a^2D_{\mu}}(\omega'^2 + 1)) d\omega' + \frac{1}{4\pi} \int_{-\infty}^{+\infty} \log(r^{\mu-m} + \frac{D_m}{a^2D_{\mu}}(\omega'^2 + 1)) d\omega' \nonumber\\
    &= -\frac{1}{4\pi} \int_{-\infty}^{+\infty}\log(\omega'^2 + 1)d\omega' + \frac{1}{4 \pi} \int_{-\infty}^{+\infty} \log \left ( \omega'^2 + 1 + \frac{a^2D_{\mu}}{D_m}r^{\mu-m} \right ) d\omega' \nonumber\\
    &= -\frac{1}{2} - C + \frac{1}{2} \sqrt{1 + \frac{a^2D_{\mu}}{D_m}r^{\mu-m}} + C = \frac{1}{2}\left (
    \sqrt{1 + \frac{a^2D_{\mu}}{D_m}r^{\mu-m}} - 1
    \right ) \; .
\end{align}
Since the divergent constant is the same in both integrals, it cancels out and $\hat R_{m\mu}$ diverges as $r^{(\mu-m)/2}$, exactly as the width $\Delta_{m\mu}$ of the coherence.

Inspecting the power spectra $S_{\mu\mu}(\omega')$ and $S_{mm}(\omega')$ we can notice that the feedback link transmits the slow fluctuations of its source. The slow fluctuations of $x_{\mu}$ are transmitted almost unchanged to $x_m$ while the intrinsic fast dynamics of $x_m$ remains superimposed. In fact, for $\omega'\ll r^{\mu-m}$:
 \begin{equation}
 S_{mm}(\omega') \approx S^{(0)}_{m}(\omega') + a^2S^{(0)}_{\mu}(\omega')
 \end{equation}
 i.e. the power spectrum of $x_m$ tends to be the sum of the natural spectra of the two dofs. The transmission of the slow fluctuations over a growing coherence width qualitatively justifies the divergence of the rescaled mutual information rate.

\subsection{Feedback link starting from $m < \mu$}
\label{Feedback_alto}
\noindent We now consider the feedback link $x_m \Rightarrow x_l$, with $l < m < \mu$. Unlike the previous case, both interacting variables become asymptotically fast with respect to the reference timescale $\tau_{\mu}$ as $r \to \infty$. The corresponding equations of motion are
 \begin{align}
    \dot x_l &= -r^{\mu-l}x_l + a r^{\mu-l}x_m + \sqrt{2 D_l r^{\mu-l}}\xi_l(t') \nonumber\\
    \dot x_m &= -r^{\mu-m}x_m + \sqrt{2D_m r^{\mu-m}}\xi_m(t')
\end{align}
and, repeating the calculation of the previous subsections, one obtains the coherence
\begin{equation}
\phi_{lm}(\omega') = \frac{a^2 D_m r^{m+2 \mu}}{a^2 D_m r^{m+2 \mu}+D_l r^l \left(\omega'^2 r^{2 m}+r^{2 \mu}\right)}
\end{equation}
so that the dimensionless mutual information rate diverges as
\begin{equation}
\hat R^{(\mu)}_{lm} \sim r^{(2\mu - m -l)/2} \; .
\end{equation}
As in the previous feedback-link case, the divergence originates from the fact that the coherence remains of order unity over a frequency interval whose width increases with the timescale separation.

\subsection{Direct link following a feedback link}
\noindent So far, we have shown that a feedback link modifies the spectrum of the receiving node by introducing the slow spectral component of the source. Hence, a subsequent direct link that follows a feedback link acts on both the natural spectrum of the fast variable and the slow component inherited from the feedback connection.

\subsubsection{Feedback link enclosing a subsequent direct link}
\label{sec:enclosed_motif}
\noindent Suppose that we have a feedback link starting from dof $x_{\mu}$ and ending in dof $x_l$ and then a subsequent direct edge from $x_l$ to $x_m$, $x_{\mu} \Rightarrow x_l \Rightarrow x_m$, where $l < m < \mu$. We investigate how the low-pass filtering action of the direct link is modified by the preceding feedback connection. The equations of motion are
\begin{align}
    \dot x_l &= -r^{\mu-l}x_l + ar^{\mu-l} x_{\mu} + \sqrt{2D_lr^{\mu-l}}\xi_l(t') \nonumber\\
    \dot x_m &= -r^{\mu-m}x_m   + b r^{\mu-m}x_l +\sqrt {2D_mr^{\mu-m}}\xi_m(t') \nonumber\\
    \dot x_{\mu} &= -x_{\mu}   + \sqrt {2D_{\mu}}\xi_{\mu}(t')
\end{align}
and we have
\begin{align}
    S_{ll}(\omega') &= |H_{l\mu}(\omega')|^2S^{(0)}_{\mu}(\omega') + S^{(0)}_l(\omega') \nonumber\\
    S_{mm}(\omega') &= |H_{ml}(\omega')|^2S_{ll}(\omega') + S^{(0)}_m(\omega') \nonumber\\
    S_{\mu\mu}(\omega') &= S^{(0)}_{\mu}(\omega')
\end{align}
which shows that the feedback link makes $S_{ll}$ contain both the intrinsic fast spectrum $S^{(0)}_l$ and the slow spectrum $S^{(0)}_{\mu}$ inherited from $x_{\mu}$.
The coherence between nodes $l$ and $m$ then becomes
\begin{equation} \phi_{lm}(\omega') = |H_{ml}(\omega')|^2 \frac{S_{ll}(\omega')}{S_{mm}(\omega')} =\frac{b^2 r^{2 \mu} \left(D_l \left(\omega'^2+1\right) r^l+a^2D_{\mu} r^\mu\right)}{\left(\omega'^2+1\right) \left(b^2D_l r^{l+2 \mu}+D_m r^m \left(\omega'^2 r^{2 l}+r^{2 \mu}\right)\right)+a^2b^2D_{\mu} r^{3 \mu}} \end{equation}
which, in the low-frequency band $\omega'=\mathcal O(r^{(\mu-m)/2})$, gives
\begin{equation}
 \phi_{lm}(\omega') \approx \frac{r^{\mu-m}}{r^{\mu-m} + \frac{D_m}{a^2b^2D_{\mu}}(\omega'^2 +1)} \; .
\end{equation}
Remarkably, this low-frequency expression is identical to the coherence generated by a single feedback link from $x_{\mu}$ to $x_m$ with effective coupling strength $ab$, since the peak of the full coherence satisfies $\mathcal{A}_{lm}(r) \to 1$ in the limit $r \to \infty$ and its half-width scales as $\Delta_{lm}(r) \sim r^{(\mu-m)/2}$. This equivalence originates from the spectral decomposition of $S_{ll}$. The feedback link adds the slow spectral component $S^{(0)}_{\mu}$ to the intrinsic spectrum $S^{(0)}_l$. Since the cutoff frequency of the direct link satisfies $\omega'_c \sim r^{\mu-m} \gg 1$, the slow component is transmitted almost unaffected, whereas only the fast frequencies of $S^{(0)}_l$ are attenuated.

The intrinsic fluctuations of $x_l$ also produce a weak coherence extending up to $\omega'\sim r^{\mu-l}$. This broader contribution determines the rate
\begin{equation}
\hat R^{(\mu)}_{lm}\sim\frac{b^2D_l}{4D_m}r^{\mu-m},\qquad
\hat R_{lm}\longrightarrow\frac{b^2D_l}{4D_m}
\end{equation}
which shows that the rate of the direct pair remains finite when measured in units of $\tau_m$. An interesting property that we will use in the following is that, in the same low-frequency band, we can write
\begin{equation}
\phi_{\mu m}(\omega') \approx \phi_{lm}(\omega') \approx \frac{r^{\mu-m}}{r^{\mu-m} + \frac{D_m}{a^2b^2D_{\mu}}(\omega'^2 +1)}
\end{equation}
meaning that, effectively, dof $x_m$ ``feels'' a feedback connection $x_{\mu} \Rightarrow x_m$ whose weight is $ab$. The endpoint rate therefore scales as $\hat R_{\mu m}\sim(|ab|/2)\sqrt{D_{\mu}/D_m}\,r^{(\mu-m)/2}$.

\subsubsection{Feedback link not enclosing a subsequent direct link}
\noindent Another interesting combination is when a direct link follows a feedback link, but it is not enclosed by it. Consider the configuration: $x_m \Rightarrow x_l \Rightarrow x_{\mu}$, where $l < m <\mu$. The corresponding equations are
\begin{align}
    \dot x_l &= -r^{\mu-l}x_l + ar^{\mu-l} x_m + \sqrt{2D_lr^{\mu-l}}\xi_l(t') \nonumber\\
    \dot x_m &= -r^{\mu-m}x_m    +\sqrt {2D_mr^{\mu-m}}\xi_m(t') \nonumber\\
    \dot x_{\mu} &= -x_{\mu}   + bx_l +  \sqrt {2D_{\mu}}\xi_{\mu}(t')
\end{align}
and in the limit $r \to \infty$, below the cutoff $\omega'\sim r^{\mu-l}$ of $x_l$, we have
\begin{align}
    S_{ll}(\omega') &\approx  a^2S^{(0)}_m(\omega') + S^{(0)}_l(\omega') \nonumber\\
    S_{mm}(\omega') & \approx  S^{(0)}_m(\omega')
    \nonumber\\
    S_{\mu\mu}(\omega') &\approx  S^{(0)}_{\mu}(\omega') \; .
\end{align}
Because of the presence of the feedback link $x_m \Rightarrow x_l$, $S_{ll}$ is split into the sum of the natural spectrum $S^{(0)}_l(\omega')$ and $S^{(0)}_m(\omega')$ multiplied by the square of the edge weight, $a^2$. The filtering action of the direct link is visible in $S_{\mu\mu}$, which approaches $S^{(0)}_{\mu}(\omega')$. Indeed, both $S_l^{(0)}$ and the component inherited from $x_m$, which is still fast with respect to $\tau_{\mu}$, are attenuated. These weak transmitted components contribute to the information rate.

We are interested in the coherence between $x_l$ and $x_{\mu}$:
\begin{equation}
\phi_{l\mu}(\omega') =
\frac{
b^2 r^\mu \left(a^2 D_m r^{m+2\mu} + D_l r^l \left(r^{2\mu} + r^{2m}\omega'^2\right)\right)
}{
D_{\mu} \left(r^{2\mu} + r^{2l}\omega'^2\right)\left(r^{2\mu} + r^{2m}\omega'^2\right)
+ b^2 r^\mu \left(a^2 D_m r^{m+2\mu} + D_l r^l \left(r^{2\mu} + r^{2m}\omega'^2\right)\right)
}
\end{equation}
so, preserving again only the leading orders in $r$ of the numerator and the denominator, we find that the coherence is the sum of two Lorentzians:
\begin{equation}
\phi_{l\mu}(\omega') \approx \frac{b^2}{D_{\mu}}\left[ \frac{a^2 D_m\, r^{\mu-m}}{r^{2(\mu-m)} + \omega'^2} + \frac{D_l\, r^{\mu-l}}{r^{2(\mu-l)} + \omega'^2} \right] \; .
\end{equation}
The first term has peak $\sim r^{m-\mu}$ and width $\sim r^{\mu-m}$, exactly as a simple direct link from $m$ to $\mu$ with weight $ab$, and dominates for $\omega' \lesssim r^{\mu-m}$. The second term, which originates from the natural spectrum of $x_l$, is a lower plateau with peak $\sim r^{l-\mu}$ that extends up to $\omega' \sim r^{\mu-l}$. Both terms have a product of peak and width of order one, so that both contribute to a finite $\hat R_{l\mu}$. Using as before the fact that $-\log(1-\phi_{l\mu}(\omega')) \approx \phi_{l\mu}(\omega')$, we obtain
\begin{equation}
\hat R_{l\mu} \to \frac{b^2\left(a^2 D_m + D_l\right)}{4 D_{\mu}}, \qquad \hat R_{m\mu} \to \frac{a^2 b^2 D_m}{4 D_{\mu}},
\end{equation}
where the latter coincides with the finite value of a direct link $x_m \Rightarrow x_{\mu}$ with weight $ab$.

\subsubsection{Feedback link originating from $x_m$ with $m < \mu$}
\noindent The same phenomenology is observed also when the feedback link originates from a node faster than $x_{\mu}$. In this case, the asymptotic behavior of the coherence follows directly from the scaling laws derived for isolated feedback links originating from a node faster than $x_{\mu}$ as discussed in Sections~\ref{direct_alto} and~\ref{Feedback_alto}.

\subsection{Summary of the elementary links}
\noindent Table~\ref{tab2} summarizes the leading-order scaling of the peak height $\mathcal{A}$ and of the half-width $\Delta$ of the coherence, and of their product, for all the elementary links discussed above. For an enclosed direct link, these quantities describe the low-frequency peak discussed in Section~\ref{sec:enclosed_motif}; the broader weak contribution must also be included when computing the rate.

\begin{table}[h]
\centering
\begin{tabular}{| c || c | c | c |}
\hline
\textbf{Elementary link} & \textbf{Peak $\mathcal{A}$} & \textbf{width $\Delta$} &  $\mathcal{A} \times \Delta$  \\
\hline
\raisebox{-0.4\height}{\includegraphics[width=0.2\linewidth]{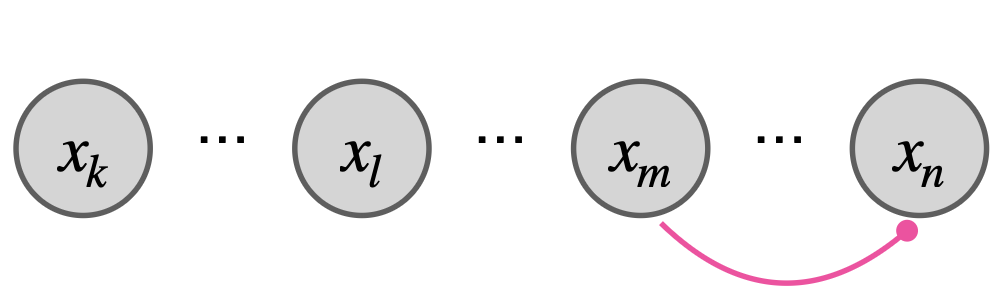}}
& $1/r^{\mu-m}$ &$r^{\mu - m}$ & $1$  \\
\hline
\raisebox{-0.4\height}{\includegraphics[width=0.2\linewidth]{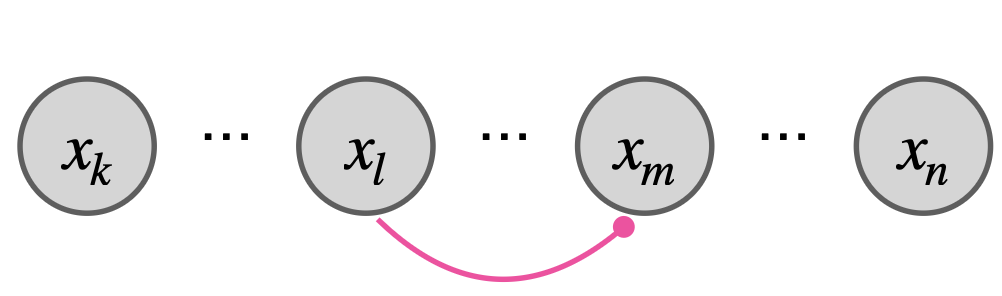}} & $1/r^{m-l}$ &$r^{\mu-l}$ & $r^{\mu-m}$  \\
\hline
\raisebox{-0.4\height}{\includegraphics[width=0.2\linewidth]{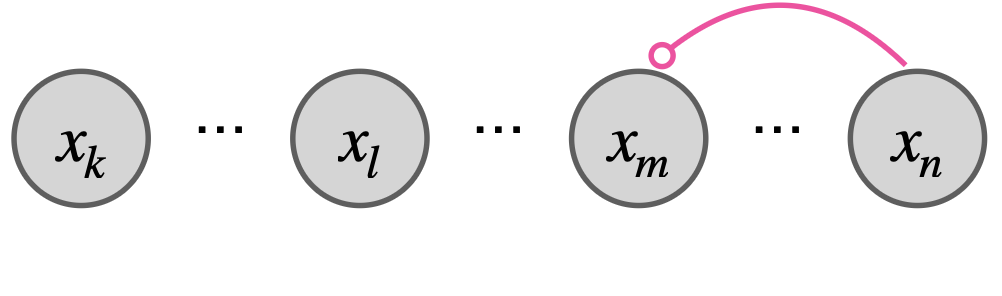}} & 1 & $r^{(\mu-m)/2}$ & $r^{(\mu-m)/2}$  \\
\hline
\raisebox{-0.4\height}{\includegraphics[width=0.2\linewidth]{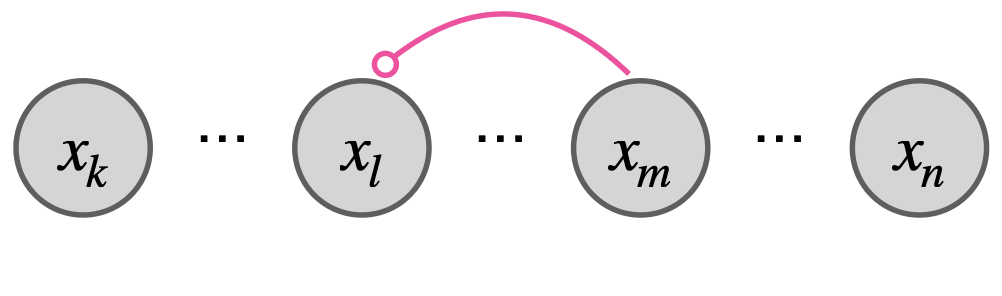}} & 1 & $r^{((\mu -m) + (\mu-l))/2}$ & $r^{((\mu -m) + (\mu-l))/2}$  \\
\hline
\raisebox{-0.4\height}{\includegraphics[width=0.2\linewidth]{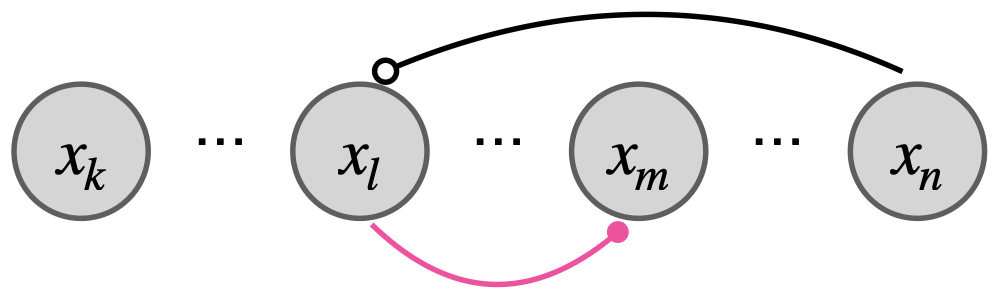}} & $1$ & $r^{(\mu-m)/2}$ & $r^{(\mu-m)/2}$
\\
\hline
\raisebox{-0.4\height}{\includegraphics[width=0.2\linewidth]{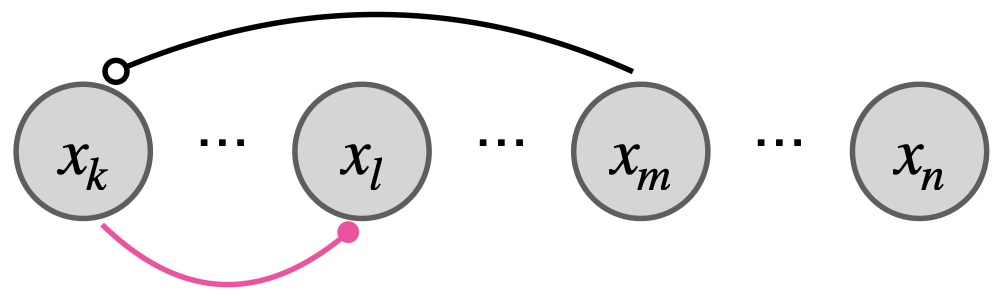}} & $1$ & $r^{((\mu-m) + (\mu - l))/2}$ & $r^{((\mu-m) + (\mu - l))/2}$
\\
\hline
\raisebox{-0.4\height}{\includegraphics[width=0.2\linewidth]{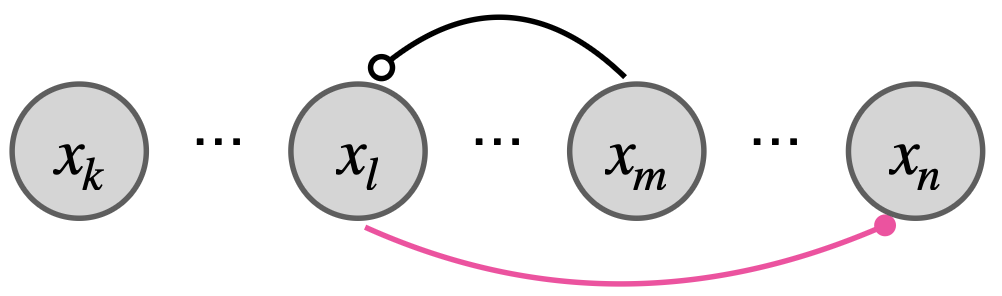}} & $1/r^{\mu-m}$ & $r^{\mu-m}$ & $1$
\\
\hline
\raisebox{-0.4\height}{\includegraphics[width=0.2\linewidth]{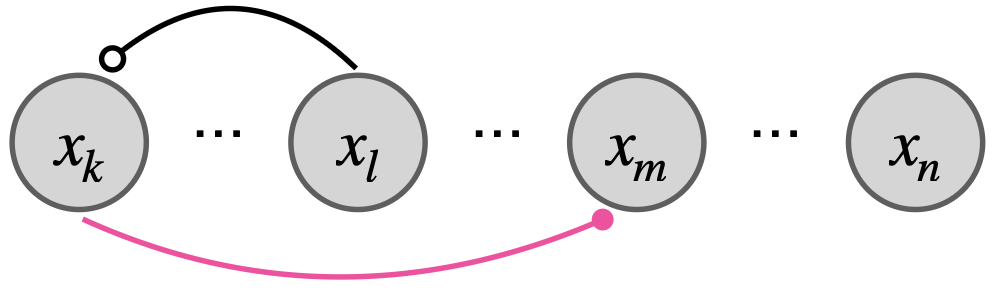}} & $1/r^{m-l}$ & $r^{\mu-l}$ & $r^{\mu-m}$
\\

\hline

\hline
\end{tabular}
\caption{The table summarizes the leading-order scaling of the coherence peak height and width for all elementary links highlighted in magenta in the first column. In the second column we report the scaling of the peak $\mathcal{A}$; in the third column there is the scaling of the width $\Delta$; in the fourth column we report the scaling of the product $\mathcal{A} \times \Delta$. We assume this order: $k < l < m < \mu$.}
\label{tab2}
\end{table}

\section{MIRMO of all three-node trees}
\label{sec:identifiability}
\noindent In this section, we apply the principles formulated above to the trees with $N = 3$ nodes $x_1, x_2, x_3$, with $\tau_1 \ll \tau_2 \ll \tau_3$, and $L = 2$ links. For these values of $N$ and $L$ there are 12 possible trees and, for each of them, we compute the rescaled rates and the MIRMO matrix, as defined in the main text. The rates can tend to zero, a finite positive value or infinity, which we encode in the labels $\mathcal L_{ij} = 0,1,2$, respectively.  

\begin{figure}[t]
    \centering
    \includegraphics[width=0.9\linewidth]{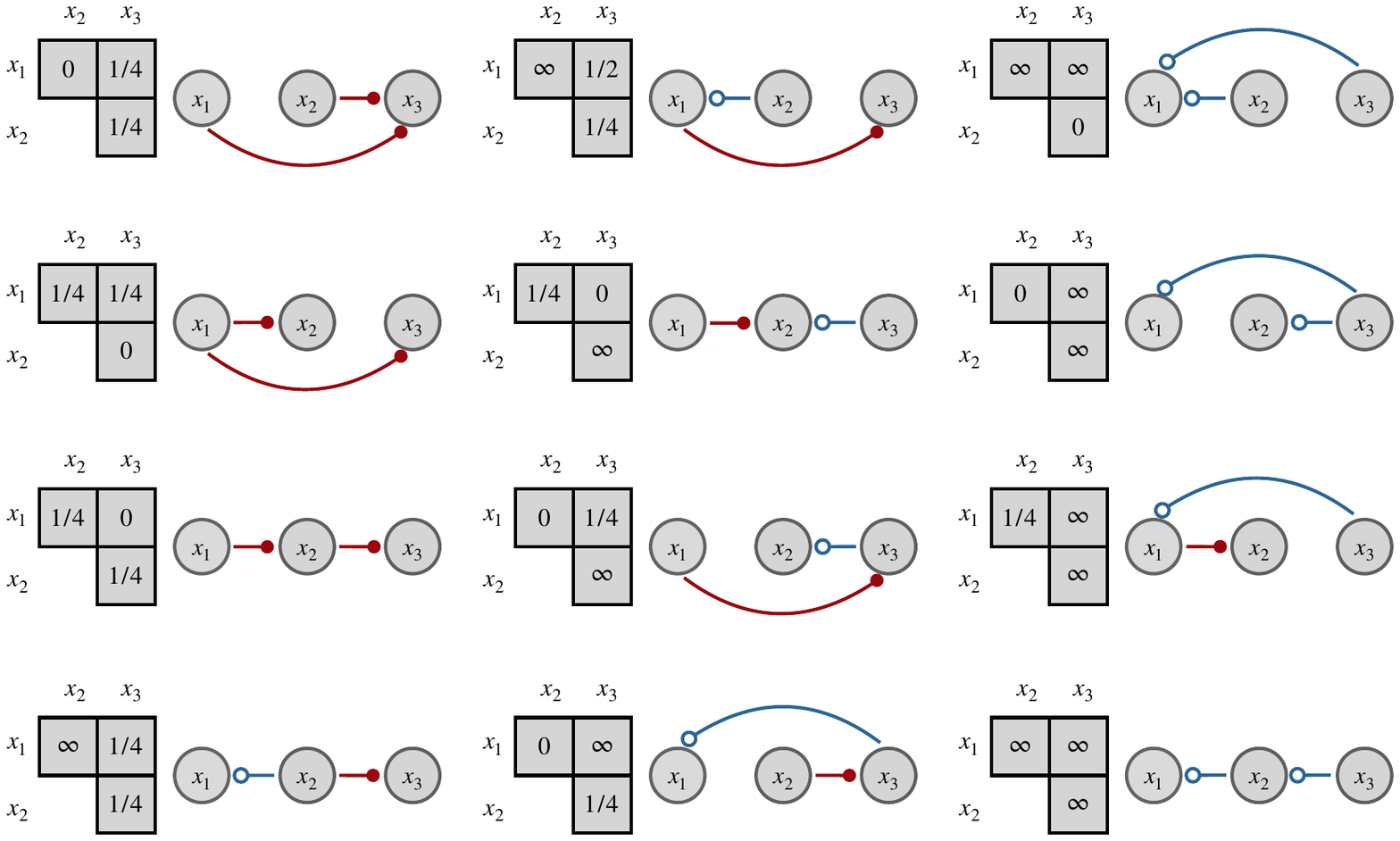}
    \caption{All the 12 trees with $N = 3$ nodes and $L = 2$ links and the corresponding rescaled rates $\hat R_{ij}$ in the timescale-separation limit $r \to \infty$. For the computation of the finite values, we assume both the coupling constants and the diffusion coefficients to be equal to unity.}
    \label{fig:treesN3L2}
\end{figure}

\begin{table}[t]
\centering
\begin{tabular}{l c c c c}
\hline
Tree & $\hat R_{12}$ & $\hat R_{23}$ & $\hat R_{13}$ & $(\mathcal L_{12},\mathcal L_{23},\mathcal L_{13})$ \\
\hline
$1\Rightarrow2,\ 2\Rightarrow3$ & $1/4$ & $1/4$ & $0$ & $(1,1,0)$\\
$1\Rightarrow2,\ 3\Rightarrow1$ & $1/4$ & $\infty$ & $\infty$ & $(1,2,2)$\\
$2\Rightarrow1,\ 1\Rightarrow3$ & $\infty$ & $1/4$ & $1/2$ & $(2,1,1)^\ast$\\
$1\Rightarrow3,\ 3\Rightarrow2$ & $0$ & $\infty$ & $1/4$ & $(0,2,1)$\\
$3\Rightarrow2,\ 2\Rightarrow1$ & $\infty$ & $\infty$ & $\infty$ & $(2,2,2)$\\
$2\Rightarrow3,\ 3\Rightarrow1$ & $0$ & $1/4$ & $\infty$ & $(0,1,2)$\\
$1\Rightarrow2,\ 3\Rightarrow2$ & $1/4$ & $\infty$ & $0$ & $(1,2,0)$\\
$1\Rightarrow3,\ 2\Rightarrow3$ & $0$ & $1/4$ & $1/4$ & $(0,1,1)$\\
$2\Rightarrow1,\ 3\Rightarrow1$ & $\infty$ & $0$ & $\infty$ & $(2,0,2)$\\
$1\Rightarrow2,\ 1\Rightarrow3$ & $1/4$ & $0$ & $1/4$ & $(1,0,1)$\\
$2\Rightarrow1,\ 2\Rightarrow3$ & $\infty$ & $1/4$ & $1/4$ & $(2,1,1)^\ast$\\
$3\Rightarrow1,\ 3\Rightarrow2$ & $0$ & $\infty$ & $\infty$ & $(0,2,2)$\\
\hline
\end{tabular}
\caption{Asymptotic rescaled rates and MIRMO labels of the 12 trees with $N=3$, with all couplings and diffusion coefficients equal to one. In a fork one dof drives two. The two trees marked with an asterisk share the same labels.}
\label{tab:trees3}
\end{table}

In Fig. \ref{fig:treesN3L2} we show for each of the graphs the corresponding MIRMO, which is also reported in Tab.~\ref{tab:trees3}. Two of them, the chain $x_2\Rightarrow x_1\Rightarrow x_3$ and the fork $x_2\Rightarrow x_1$, $x_2\Rightarrow x_3$, share the same MIRMO labels, while all the others differ from each other. Notably, the number of possible MIRMO label triplets in this case is $3^3 = 27$, which is larger than the number of possible trees. Some of the remaining triplets correspond to graphs with a single edge (for instance this MIRMO,  $\hat R_{12} \to \infty$,  $\hat R_{13}\to 0$ and $\hat R_{23} \to 0$, corresponds to the feedback link $x_2 \Rightarrow x_1$ and the isolated node $x_3$). Others correspond to graphs with $L = 3$ links. For instance, a MIRMO matrix with all off-diagonal labels equal to one is produced by the feed-forward triangle $x_1 \Rightarrow x_2$, $x_1 \Rightarrow x_3$, $x_2 \Rightarrow x_3$, in which every link is direct and not enclosed. 

Let us show explicitly that the chain and the fork share the same labels. The pair $(x_1,x_2)$ obeys the same dynamics in both trees, so that $\hat R_{12}\sim r^{1/2}$ in both cases. On the timescale $\tau_3$, $x_1$ is a noisy copy of $x_2$, since $\phi_{12}=1-\mathcal O(r^{-1})$ for $\omega'=\mathcal O(r)$. Since the noises are independent, the coherence factorizes exactly through the dof in the middle of the pathway, i.e., $\phi_{23}=\phi_{12}\,\phi_{13}$ for the chain and $\phi_{13}=\phi_{12}\,\phi_{23}$ for the fork. Denoting by $a$ the weight of $x_2\Rightarrow x_1$ and by $b$ the weight of the link into $x_3$, we find
\begin{equation}
\begin{aligned}
\text{chain:}\qquad &\hat R_{23}\to\frac{a^2b^2D_2}{4D_3},\qquad \hat R_{13}\to\frac{b^2(a^2D_2+D_1)}{4D_3};\\
\text{fork:}\qquad &\hat R_{23}\to\frac{b^2D_2}{4D_3},\qquad \hat R_{13}\to\frac{b^2D_2}{4D_3},
\end{aligned}
\end{equation}
so that both trees have labels $(2,1,1)$. Physically, a fast dof driven by a slower one acts as a relay of its driver on slower timescales, and attaching a link to the driver or to its relay changes the finite values of the rates, but not their labels.

\subsection{Distinguishing the fork and the chain by conditioning}
\label{sec:conditional_example}
\noindent Let us now compute the conditional rates for the same two trees. We keep the coupling $a$ from $x_2$ to $x_1$, while $b$ couples $x_2$ to $x_3$ in the fork and $x_1$ to $x_3$ in the chain. In terms of the physical frequency $\omega$, their Fourier-space matrices are
\begin{equation}
B_{\rm fork}=\begin{pmatrix}
1+i\omega\tau_1 & -a & 0\\
0 & 1+i\omega\tau_2 & 0\\
0 & -b & 1+i\omega\tau_3
\end{pmatrix},
\qquad
B_{\rm chain}=\begin{pmatrix}
1+i\omega\tau_1 & -a & 0\\
0 & 1+i\omega\tau_2 & 0\\
-b & 0 & 1+i\omega\tau_3
\end{pmatrix}.
\label{eq:fork_chain_B}
\end{equation}
Using Eq.~\eqref{eq:inverse_spectrum}, with $q_i=2D_i\tau_i$, we obtain the off-diagonal elements
\begin{equation}
\begin{aligned}
P_{12}^{\rm fork}=P_{12}^{\rm chain}
&=-\frac{a(1-i\omega\tau_1)}{q_1},\\
P_{13}^{\rm fork}=0,\qquad
P_{23}^{\rm fork}&=-\frac{b(1+i\omega\tau_3)}{q_3},\\
P_{13}^{\rm chain}=-\frac{b(1+i\omega\tau_3)}{q_3},\qquad
P_{23}^{\rm chain}&=0.
\end{aligned}
\label{eq:fork_chain_P}
\end{equation}
For non-zero $a$ and $b$, the diagonal elements of $P$ are positive, so that Eq.~\eqref{eq:conditional_mir} gives
\begin{equation}
\begin{aligned}
\text{fork:}\qquad &R_{12|3}>0,\qquad R_{13|2}=0,\qquad R_{23|1}>0,\\
\text{chain:}\qquad &R_{12|3}>0,\qquad R_{13|2}>0,\qquad R_{23|1}=0.
\end{aligned}
\label{eq:fork_chain_conditional_rates}
\end{equation}
In the fork, the trajectory of $x_2$ accounts for the fluctuations shared by $x_1$ and $x_3$. Once it is known, their independent noises leave no further dependence. In the chain, conditioning on the intermediate dof $x_1$ instead removes the dependence between $x_2$ and $x_3$. The conditional rates therefore distinguish the two networks at every finite $r$, as shown in Fig.~2b of the main text.

\subsection{Reconstruction from the two rates}
\label{sec:reconstruction}
\noindent The End Matter in the main text extends this construction to a fully observed tree with stationary linear Gaussian dynamics, independent noises with $D_i>0$, fixed non-zero couplings, and one directed interaction per edge. The ordinary rate detects fluctuations shared through a common ancestor, while conditioning removes dependence carried by an intermediate dof and can reveal dependence between two dofs driving a common target. Together, these properties identify a link through
\begin{equation}
x_i\text{ and }x_j\text{ are linked}
\quad\Longleftrightarrow\quad
R_{ij}>0\quad\text{and}\quad R_{ij|-ij}>0.
\label{eq:link_criterion}
\end{equation}
The rates in this test are evaluated at finite $r$. Once the links are identified, their MIRMO labels determine the directions in the timescale-separation limit: for $i<j$, $\mathcal L_{ij}=1$ gives $x_i\Rightarrow x_j$, while $\mathcal L_{ij}=2$ gives $x_j\Rightarrow x_i$. The result for the label of a link embedded in a tree is proved in the End Matter. The same reconstruction applies to forests, since dofs in different connected components are independent.

\section{Simulations}
\label{sec:simulations}

\subsection{Dynamics and parameters}
\noindent For the figures in the main text, we simulated Eq.~\eqref{eq:langevin_system} with $f_{ij}(x_j)=a_{ij}f(x_j)$. The stochastic equations were integrated using the Euler-Maruyama method with time step $dt=0.1$ and trajectories up to $10^8$ time steps. In all cases, we set $D_i=1$.

For linear couplings, $f(x)=x$, we set $\tau_1=1$ and used $r=\tau_3/\tau_2=\tau_2/\tau_1=2,5,10,20$. Figure~2d of the main text shows the trajectory-length dependence for $x_3\Rightarrow x_1\Rightarrow x_2$. Its finite-$r$ reference spectra follow from Section~\ref{sec:enclosed_motif} with $(\mu,l,m)=(3,1,2)$.

To study nonlinear couplings, we used $f(x)=\beta^{-1}\tanh(\beta x)$ with $\beta=0.1,0.5,1,2$, and $f(x)=x-\epsilon x^3$ with $\epsilon=0.001,0.01,0.1,0.2$, at $r=10$. The hyperbolic tangent approaches the linear coupling when $|\beta x|\ll1$. For these trajectories, the coherence integral gives a Gaussian spectral approximation to the mutual information rate.

\subsection{Spectral calculation of the rates}
\label{sec:rate_estimation}
\noindent The power spectra and cross-spectra were estimated using the Welch method. The ordinary rates were obtained by computing the coherence and integrating Eq.~\eqref{eq:mir_gauss}. For a pair $A=\{i,j\}$ and conditioning set $C$, the conditional spectrum can be computed from the same spectral matrix as
\begin{equation}
S_{A|C}(\omega)=S_{AA}(\omega)-S_{AC}(\omega)S_{CC}(\omega)^{-1}S_{CA}(\omega).
\label{eq:conditional_spectrum_blocks}
\end{equation}
The coherence conditioned on $C$ is
\begin{equation}
\phi_{ij|C}(\omega)=\frac{|[S_{A|C}(\omega)]_{ij}|^2}{[S_{A|C}(\omega)]_{ii}[S_{A|C}(\omega)]_{jj}}.
\label{eq:conditional_coherence_blocks}
\end{equation}
For $C=-ij$, $S_{A|C}=(P_{AA})^{-1}$ this expression gives the partial coherence defined in Section~\ref{sec:notation}. For an empty conditioning set it reduces to the ordinary coherence. For real-valued trajectories, the coherence is even in frequency, so a quadrature estimate on a nonnegative angular-frequency grid $\omega_k$ has the form
\begin{equation}
\hat R^{\mathrm{est}}_{ij|C}=-\frac{\tau_{\max(i,j)}}{2\pi}\sum_k w_k\log\left[1-\phi^{\mathrm{est}}_{ij|C}(\omega_k)\right],
\label{eq:rate_quadrature}
\end{equation}
where $w_k$ are the integration weights in angular frequency. Increasing the trajectory length at fixed $r$ probes convergence to a finite rate; the MIRMO labels describe its dependence on $r$.

\end{document}